%% file: main.tex
\documentclass[%
  reprint,
nofootinbib,
nobibnotes,
 amsmath,amssymb,
pra,
]{revtex4-1}

\usepackage{graphicx}
\usepackage{dcolumn}
\usepackage{bm}
\usepackage{hyperref}
\usepackage{mathtools} 
\usepackage{verbatim}

\usepackage{amsmath}
\usepackage{amsthm}
\usepackage{physics}
\usepackage[oldvoltagedirection]{circuitikz}
\usepackage{tabularx}

\usepackage{pgfplots}
\usepgfplotslibrary{external}
\pgfplotsset{compat=1.9}
\usepackage{tikz}
\usetikzlibrary{matrix}

\pgfplotsset{
    myplotstyle/.style={
    legend style={draw=none, font=\small},
    legend cell align=left,
    legend pos=north east,
    ylabel style={align=center, font=\bfseries\boldmath},
    xlabel style={align=center, font=\bfseries\boldmath},
    x tick label style={font=\bfseries\boldmath},
    y tick label style={font=\bfseries\boldmath},
    scaled ticks=false,
    every axis plot/.append style={thick},
    },
}

\newcommand{\Z}{\mathbb{Z}}

\newcommand{\ad}{\hat{a}^{\dagger}}

\newcommand{\Lv}{\mathcal{L}}

\allowdisplaybreaks

\begin{document}


\title{Characterising Polariton States in Non-Dispersive Regime of Circuit Quantum Electrodynamics}

\author{Arvind Mamgain}
\author{Samarth Hawaldar}
\author{Athreya Shankar}
\author{Baladitya Suri}
\affiliation{Department of Instrumentation and Applied Physics, Indian Institute of Science, Bengaluru, India}%
\date{\today}
\begin{abstract}
A superconducting qubit coupled to a read-out resonator is currently the building block of multiple quantum computing as well as quantum optics experiments. 
A typical qubit-resonator system is coupled in the dispersive regime, where the detuning between qubit and resonator is much greater than the coupling between them. In this work, we fabricated and measured a superconducting transmon-resonator system in the non-dispersive regime. The dressed states formed by the mixing of the bare qubit and resonator states can be further mixed by applying a drive on the qubit, leading to the formation of polariton states. We report experimental studies of transitions between polariton states at varying driving powers and frequencies and show how the non-dispersive coupling of the higher levels of the qubit-resonator system modifies the polariton eigenstates and the corresponding transition frequencies. We also report close agreement with numerical results obtained from a driven Jaynes–Cummings Model beyond the dispersive regime.
\end{abstract}

\maketitle

\section{\label{sec:Introduction}Introduction}

Circuit quantum electrodynamics (cQED) is an excellent testbed for the study of various light-matter, and matter-matter interaction phenomena \cite{wallraff2004strong,blais2004cavity, you_atomic_2011,Gu2017a}. A basic building block of cQED architectures consists of a microwave resonator coupled to an artificial atom, both of which are realized on a chip using superconducting microwave circuits. The atom-resonator unit forms a multi-level quantum system which can be controlled by applying microwave drives. Besides forming the basis for superconducting qubits for use in quantum computing, cQED systems are also attractive from a quantum optics perspective 
because they can be tuned into regimes beyond what is easily achievable, or even feasible, with natural atoms \cite{schoelkopf2008wiring, you_atomic_2011}. Examples of such phenomena explored via cQED include the implementation of strong \cite{wallraff2004strong,schuster_ac_2005}, ultra-strong \cite{niemczyk2010circuit,forn-diaz_ultrastrong_2017}, and deep strong \cite{ yoshihara2017superconducting} regimes of atom-cavity (resonator) coupling, probing the photon-number nonlinearity of Jaynes-Cummings (JC) systems \cite{fink2008climbing, bishop2009nonlinear,suri2015nonlinear}, and single atom lasing \cite{astafiev2007single}. The JC nonlinearity induced by a superconducting qubit in a resonator has also been used to implement a high-fidelity read-out \cite{reed2010high,bishop2010response,boissonneault2010improved}

Additional tunability of cQED systems can be introduced via the application of drive fields. By driving the $\Xi$ (ladder) type transitions in an artificial atom, Mollow triplets \cite{Baur2009}, Autler-Townes Splitting (ATS) \cite{sillanpaa2009autler, Baur2009,Peng2018}, and electromagnetically induced transparency (EIT) \cite{ian2010tunable,murali2004probing,sun2014electromagnetically} have been probed and a single-photon router \cite{hoi2011demonstration} has been implemented. Additionally, a $\Lambda$-type level structure has been realised in a  qubit-resonator system using drives on both qubit-like and resonator-like transitions \cite{suri2013observation,Novikov2016}. These $\Lambda$ systems have been used to measure the coherence of a dark state \cite{Novikov2016}, the coherence of high-Q resonators by probing EIT with a sideband drive on the atom \cite{Ann2020}, and to detect single microwave photons \cite{inomata2016single}.

In particular, ``polariton" states, which are superpositions of qubit-resonator dressed states caused by additional drive fields, have been studied in~\cite{Gu2016, Long2018a, Koshino2013, Koshino2013b, Inomata2014b}. The tunability of frequencies and decay rates of transitions between polariton states has been used to implement an impedance-matched $\Lambda$-system \cite{Koshino2013}, which has been subsequently used for down-conversion and detection of microwave photons \cite{Koshino2013b,Inomata2014b,inomata2016single}. Polariton states have also been proposed for implementing a two-qubit gate between a superconducting and a flying qubit in a coplanar waveguide \cite{Koshino2017}.

In this article, we report spectroscopic measurements of polariton states formed by driving a transmon artificial atom \cite{koch2007charge} that is in turn, coupled to a lumped-element resonator. In particular, we engineer our system in a ``non-dispersive"  regime, where the coupling ($g_1$) of the first-to-second excited state transition of the transmon to the resonator mode is comparable to the detuning ($\Delta_1$) between the resonator and the transition frequencies.  This is in contrast to previous works \cite{Inomata2014b, Long2018a, PhysRevLett.124.070401} in which polariton states were observed in the ``dispersive" regime, where the coupling between the transmon and the resonator is considerably smaller compared to the detuning for all relevant transitions of the transmon. By means of eigenmode analysis and master equation calculations, we simulate the polariton transitions and observe close agreement with experiment. We also explain the qualitative differences between polariton transitions in the dispersive and non-dispersive regimes using perturbative calculations.


Previous studies on polariton states have focused on the limiting cases of the dispersive regime \cite{PhysRevLett.124.070401} and the fully non-dispersive regime of the $\ket{g} \rightarrow \ket{e}$ transition \cite{Kockum_2013}. In contrast, we work in the non-dispersive regime of the $\ket{e} \rightarrow \ket{f}$ transition while the $\ket{g} \rightarrow \ket{e}$ transition is dispersive. As we will demonstrate, this leads us to an interesting regime which enables us to explain the observed spectral features qualitatively within the dispersive approximation, while at the same time requiring us to incorporate the non-dispersive effects in our calculation for good quantitative agreement with the data.

The paper is organized as follows. In section \ref{sec:Theoretical model}, we describe the Hamiltonian of a transmon-resonator system including a drive field on the transmon. We discuss how our device differs from typical dispersively coupled transmon-resonator systems and introduce the concept of polariton states. In section \ref{sec:Experimental setup}, we describe the experimental setup used to characterise the device we fabricated. In section \ref{sec:Observations and results}, we discuss the experimentally measured polariton spectra obtained for varying powers and frequencies of the drive tone. We demonstrate very close agreement of the measured spectra with results obtained from an eigenmode analysis as well as full master equation simulations. In section \ref{sec:Comparison with dispersive case}, we qualitatively explain the behaviour of polariton transitions in the non-dispersive regime using perturbation theory, and compare it to the usual dispersive case. We conclude with a summary.   

\section{\label{sec:Theoretical model}Theoretical model}

\begin{figure}[tb]
  \includegraphics[width=0.8\linewidth]{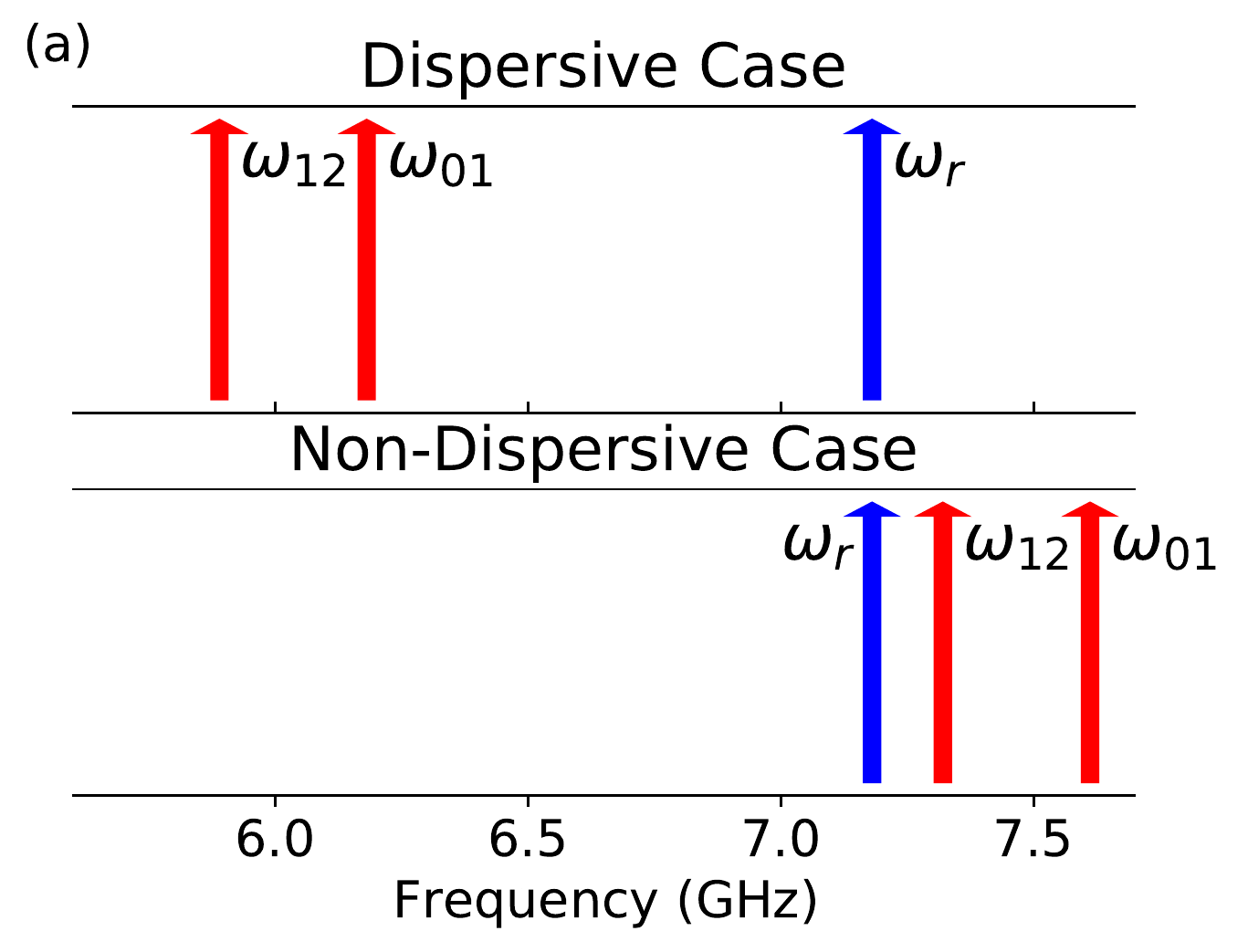}
\includegraphics[width=\columnwidth]{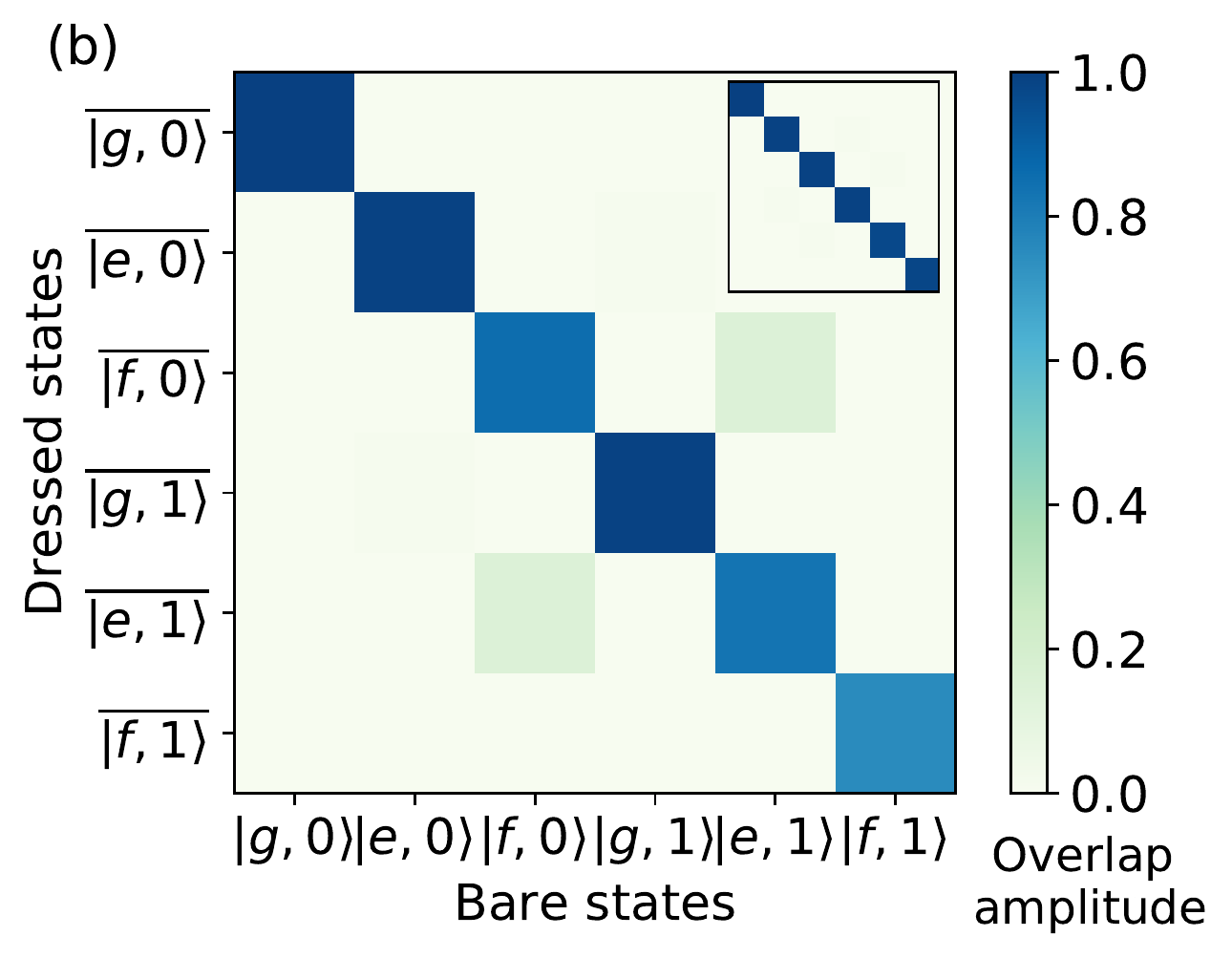}
  \caption{(a) Frequency of bare resonator and qubit for non-dispersive and dispersive case, (b) Plot showing the overlap amplitude of the dressed states with the bare states in the non-dispersive regime($g_1/\Delta_1 = 0.47$) inset in the plot show the case of the dispersive limit ($g_0/\Delta_0 = -0.108$, $g_1/\Delta_1 = -0.091$)}
  \label{disp_vs_nondisp}
\end{figure}

We consider a system consisting of a fixed-frequency transmon coupled to a resonator, which can be described by the generalised Jaynes-Cummings Hamiltonian~\cite{koch2007charge}
\begin{multline}
    \hat{H} =\hbar\omega_r \hat{a}^{\dagger}\hat{a}  + \hbar \sum_{j}\omega_j\ket{j}\bra{j} + \hbar g_0(\hat{a}^{\dagger}\hat{b} + \hat{a}\hat{b}^\dagger ).
    \label{eqn:ham_jc}
\end{multline}
Here, $\omega_r$ is the bare frequency of the resonator, $\hbar\omega_j$ is the energy of the $j^\text{th}$ excited state ($\ket{j}$) of the transmon and $g_0$ is the transmon-resonator coupling strength. The excitation (de-excitation) of the transmon and the resonator are respectively described by the creation (annihilation) operators $\hat{b}^\dag$ ($\hat{b}$) and $\hat{a}^\dag$ ($\hat{a}$). In writing Eq.~(\ref{eqn:ham_jc}), we have used the rotating-wave approximation (RWA) to neglect the rapidly oscillating terms. Furthermore, we have neglected interactions that lead to exchange of multiple excitations between the transmon and resonator, which is a valid approximation in the transmon regime~\cite{koch2007charge}. In our discussion, we will use the labels $g,e,f$ to refer to the three lowest transmon levels with $j=0,1,2$.

In our work, we introduce an additional drive on the transmon that we call the ``coupler" drive. In the presence of the coupler drive, the system Hamiltonian can be written in the rotating frame of the drive under the RWA as 
\begin{multline}\label{eq of Ham without probe}
   \hat{H}_\text{rot} = \hbar\delta_r \hat{a}^{\dagger}\hat{a}  + \hbar\sum_{n}\delta_n\ket{n}\bra{n}~\\~+~\hbar g_0(\hat{a}^{\dagger}\hat{b} + \hat{a}\hat{b}^\dagger) + \hbar\Omega_d (\hat{b} + \hat{b}^{\dagger}).
\end{multline}
Here, the drive has frequency $\omega_d$, and Rabi frequency $\Omega_d$, and we have introduced detunings $\delta_n = \omega_{n} - n\omega_d$, and $\delta_r = \omega_r - \omega_d$.

Typical transmon-resonator systems operate in the so-called dispersive regime, where the coupling strength is small compared to the detuning between the transmon transition frequencies and the resonator frequency. This condition can be expressed as $g_j/\Delta_j \ll 1$ for all energy levels $j$, where $g_j \approx g_0\sqrt{j+1}$ and $\Delta_j = \omega_{j,j+1} - \omega_r$ is the detuning of the resonator from the $\ket{j}~\rightarrow~\ket{j+1}$ transition with corresponding frequency $\omega_{j,j+1} = \omega_{j+1} - \omega_j$. The negative anharmonicity of the transmon implies that $\omega_{j,j+1} < \omega_{j-1,j}$.
For a transmon, the dispersive regime is usually realised by designing $\omega_{01}<\omega_r$ (see Fig.~\ref{disp_vs_nondisp}(a)).

However, in the case of our device, the transmon frequency is greater than the resonator frequency, such that the frequency for the $\ket{e} \rightarrow \ket{f}$ transition of the transmon falls near the resonator frequency, as shown in Fig.~\ref{disp_vs_nondisp}(a).
This leads to $g_1/\Delta_1 \approx 0.47$, for which the dispersive approximation does not hold, leading to strong modifications in the nature of the dressed states of the system. In the discussion that follows, we first discuss the properties of the dressed states in the dispersive regime and then explain how these properties change in the non-dispersive case.

\subsection{Dressed states in dispersive regime}

\subsubsection{\label{subsec:singlydressed}Singly dressed states}

\begin{figure*}[tb]
  \centering
  \includegraphics[width=\textwidth]{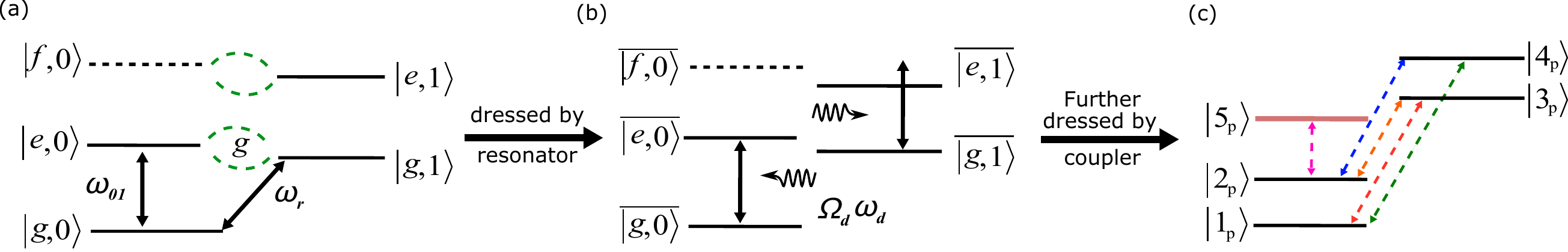}
  \caption{ Energy level diagrams showing (a) bare states of a qubit-resonator (b) dressed states of qubit-resonator showing the coupler drive (c) 5 polariton states formed for a coupler drive resonant with $\ket{g,0}$ to $\ket{e,0}$ transition with dashed arrows showing transitions between them}
  \label{Energy level diagram}
\end{figure*}

The eigenstates of the undriven transmon-resonator Hamiltonian~(\ref{eqn:ham_jc}) are the ``singly" dressed states formed by the mixing of the bare transmon and resonator eigenstates, see Fig.~\ref{Energy level diagram}(a) and (b). In the dispersive regime, the transmon-resonator coupling essentially serves to modify the bare transmon and resonator frequencies, but does not cause significant mixing of the corresponding eigenstates. As a result, the dressed states are, to very good approximation, just the bare eigenstates. This is numerically demonstrated in the inset of Fig.~\ref{disp_vs_nondisp}(b), where we plot the contribution of bare states to the different dressed states in the dispersive regime. 

The dressed states involving the lowest two transmon levels can be written in the dispersive regime as
\begin{align}
     \overline{\ket{g,n}} = \ket{g,n} - \frac{g_0\sqrt{n}}{\Delta_0}\ket{e,n-1}\nonumber\\
     \overline{\ket{e,n-1}} = \ket{e,n-1} + \frac{g_0\sqrt{n}}{\Delta_0}\ket{g,n}.
\end{align}
Here, $\overline{\ket{g,n}}$ and $\overline{\ket{e,n-1}}$ are the first and second dressed states with $n$ total excitations in the coupled transmon-resonator system. In the basis given by $\overline{\ket{g,n}}, \overline{\ket{e,n-1}}$ with $n=0,1,\ldots$, the transmon-resonator Hamiltonian in the dispersive regime can be approximated as
\begin{align}
    \hat{H}_\text{disp} = \hbar\omega_r'\hat{a}^{\dagger}\hat{a} + \hbar\omega_{01}' \hat{\sigma}_z/2 + \hbar\chi\hat{\sigma}_z\hat{a}^{\dagger}\hat{a}. 
    \label{eqn:ham_disp}
\end{align}
Defining $\chi_{j,j+1} \equiv g_{j}^2/\Delta_j$, the frequencies appearing in Eq.~(\ref{eqn:ham_disp}) are given by $  \omega_r' = \omega_r - \chi_{12}/2$, $\omega_{01}' = \omega_{01} + \chi_{01}$ and
$\chi = \chi_{01} - \chi_{12}/2$~\cite{koch2007charge}.


\subsubsection{\label{subsec:doublydressed}Doubly dressed states: Polariton states}

The inclusion of a coupler drive in the Hamiltonian~(\ref{eq of Ham without probe}) serves to introduce further structure to the levels. This drive further dresses the states, and the resultant ``doubly" dressed states are called polariton states [see Fig.~\ref{Energy level diagram}(b) and (c)]~\cite{Gu2016}. 

In the dispersive regime, for weak to moderate drive powers (i.e. $\Omega_d$), four polariton states are formed by mixing the four lowest energy singly dressed states as~\cite{Gu2016}   
\begin{equation}\label{eq:Dispersive Two-Level Polariton Eqns}
    \mqty(\ket{1_p}\\\ket{2_p}\\\ket{3_p}\\\ket{4_p}) = 
    \scalebox{1.5}{$\mqty(R_{\frac{\theta_0}{2}} & 0 \\0 & R_{\frac{\theta_1}{2}}\\)$}
    \mqty(\overline{\ket{g,0}}\\ \overline{\ket{e,0}}\\ \overline{\ket{e,1}}\\ \overline{\ket{g,1}}),
\end{equation}
where
\begin{equation*}\label{eq:Rotation matrix}
    R_{\theta} = \mqty(\cos\theta & -\sin\theta\\
    \sin\theta & \cos\theta\\)
\end{equation*}
and $\tan \theta_{0(1)} = 2\Omega_d/(\omega'_{ge,0(1)} - \omega_d)$ where $\omega'_{ge,n}$ is the frequency of the $\overline{\ket{g,n}}\rightarrow\overline{\ket{e,n}}$ transition.

\subsubsection{Polaritonic transitions}

In our experiment, we probe the transition frequencies between the various polariton states by introducing a weak probe that induces resonator-like transitions, namely $\ket{1_p}\rightarrow\ket{3_p}$, $\ket{1_p}\rightarrow\ket{4_p}$, $\ket{2_p}\rightarrow\ket{3_p}$, and $\ket{2_p}\rightarrow\ket{4_p}$. The corresponding contribution to the system Hamiltonian from the probe is given by $H_{probe} = \Omega_p (\hat{a} e^{i \omega_p t} + \ad e^{-i \omega_p t})$. 

As a point of reference, we present the polaritonic transition frequencies in the dispersive regime below, before discussing the modifications in the non-dispersive regime. When the coupler drive $\omega_d$ is resonant with $\omega'_{ge,0}$, the transition frequency of the qubit when there are zero photons in the resonator, the transition frequencies between polariton states, $\omega_{ij,p}$ (corresponding to the frequency of the $\ket{i_p} \to \ket{j_p}$ transition) are given by  
\begin{align}\label{disp_transition_freq}
    \omega_{13,p} &= \omega'_r - (\sqrt{\chi^2 + \Omega_d^2} - \Omega_d),\nonumber\\
    \omega_{14,p} &= \omega'_r + (\sqrt{\chi^2 + \Omega_d^2} + \Omega_d),\nonumber\\
    \omega_{23,p} &= \omega'_r - (\sqrt{\chi^2 + \Omega_d^2} + \Omega_d),\nonumber\\
    \omega_{24,p} &= \omega'_r + (\sqrt{\chi^2 + \Omega_d^2} - \Omega_d).
\end{align}

\subsection{Dressed states in the non-dispersive regime}

In the case of non-dispersive coupling between the transmon and the resonator, no simple closed-form expressions exist for the singly dressed states, and hence we turn to numerical inquiries to find the eigenenergies and eigenstates of the system. In Fig.~\ref{disp_vs_nondisp}(b), we plot the contribution of bare (uncoupled) transmon-resonator eigenstates to the dressed states for the parameters of our device.
In contrast to the dispersive case (inset), we observe a substantial contribution of the higher excited state $\ket{f,0}$ to the dressed state $\overline{\ket{e,1}}$.

Similarly, in the case of the doubly dressed or polariton states, we rely on numerical calculations, more specifically eigenmode simulations (Appendix~\ref{Appendix_EigenmodeSimulation}) to analyze polaritonic transitions. Experimentally observed transitions between these polariton states and comparisons with theoretically predicted transition frequencies and intensities will be discussed in detail in Section~\ref{sec:Observations and results}. For the latter, we note that the intensity of an observed transition between states $\ket{\alpha},\ket{\beta}$ is proportional to 
\begin{equation}\label{eq:IntensityTrans}
    I_{\ket{\alpha}\leftrightarrow \ket{\beta}} \propto (P_\alpha + P_\beta) \abs{\mel{\alpha}{\hat{a}}{\beta}}^2,
\end{equation}
where $P_\alpha, P_\beta$ are the steady-state occupations of the states $\ket{\alpha}$ and $\ket{\beta}$ and $\abs{\mel{\alpha}{\hat{a}}{\beta}}$ is the probe-induced transition matrix element.

Finally, we note that in non-dispersive systems, the singly dressed states are sometimes called polariton states even in the absence of an external drive. However, for clarity, in this work we reserve the term polariton states for the doubly dressed states obtained under external driving.

\section{\label{sec:Experimental setup}Experimental setup} \begin{figure}[!tb]
  \centering
  \includegraphics[trim={0cm 0cm 0cm 0cm},clip,width=0.45\textwidth]{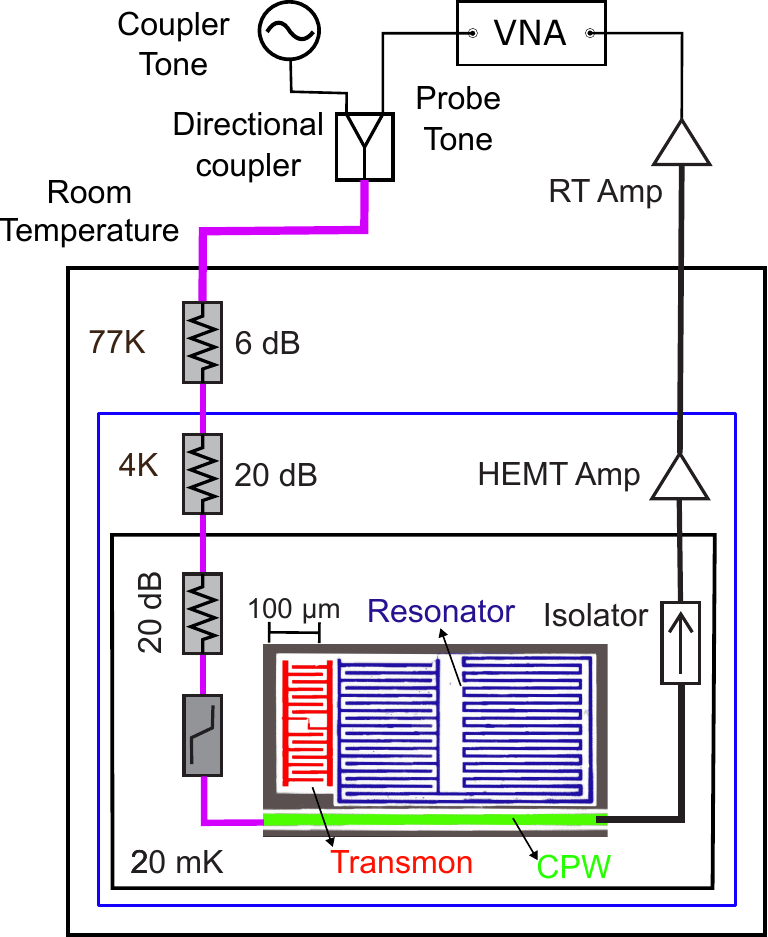}
  \caption{Schematic diagram of the measurement setup. Figure shows a false-colored SEM image of the measured device showing aluminum pattered transmon(red), resonator(blue), centre conductor of CPW(green) and ground plane(black) on silicon(white) substrate.}
  \label{Schematics}
\end{figure}

A schematic of the measurement setup is shown in Fig.~\ref{Schematics}. A fixed-frequency transmon capacitively coupled to a lumped-element resonator is mounted at the $15\,$mK stage of a dilution refrigerator. The resonator, in turn, is coupled to a coplanar waveguide transmission line (CPW) to control and probe the device. The probe and coupler tones used in the spectroscopic measurements of polariton transitions are combined using a directional coupler and input to the CPW of the device, as shown in Fig.~\ref{Schematics}. The transmitted signal coming out of the output port is then measured with the VNA after amplifying it using a high electron-mobility transistor (HEMT) amplifier at the $4\,$K stage followed by room temperature (RT) amplifiers.

All the experimentally measured and derived parameters of the device are given in Table~\ref{tab:DeviceParams}. The parameters $\omega_{01},g$ and $\alpha$ were extracted by tuning them and visually inspecting the overlay of the eigenmode simulation plots over the experimental data to get the best agreement in both Figs.~\ref{fig: experimental data - crossplot}(b) and~\ref{fig: experimental data - xplot}(b). The dressed frequencies of the qubit and resonator were extracted from qubit and resonator spectroscopy and numerically verified by diagonalizing the Hamiltonian in Eq.~(\ref{eqn:ham_jc}) using the aforementioned three parameters. The bare resonator frequency was obtained by performing resonator spectroscopy at very high probe powers. Time-domain measurements were performed by applying shaped pulses on the qubit followed by readout using high power readout methods~\cite{reed2010high} in order to obtain the coherence times ($T_1$ and $T_{\phi}$) of the qubit. Further details regarding the time-domain measurements can be found in Appendix~\ref{sec: Time domain experiment setup}.

\begin{table}[h]
    \centering
    \begin{tabularx}{0.45\textwidth}{|X|c|c|}
     \hline
Parameter & Symbol & Value \\
 \hline
Bare resonator frequency & $\omega_r/2\pi$ & 7.180 GHz \\
 \hline
 Bare qubit frequency & $\omega_{01}/2\pi$ & 7.611 GHz \\
 \hline
 Dressed resonator frequency & $\omega_r'/2\pi$ & 7.1665 GHz \\
 \hline
 Coupling & $g/2\pi$ & 46.57 MHz \\
 \hline
Anharmonicity & $\alpha/2\pi$ & -291.4 MHz \\
 \hline
 Dressed qubit frequency with $0$ photons in resonator & 
$\omega'_{ge,0}/2\pi$ & 7.616 GHz\\
 \hline
 Dressed qubit frequency with $1$ photon in resonator &$\omega'_{ge,1}/2\pi$ & 7.599 GHz\\
  \hline
 Dressed resonator frequency with qubit in ground state &$\omega'_{r,g}/2\pi$ & 7.175 GHz\\
  \hline
 Dressed resonator frequency with qubit in excited state &$\omega'_{r,e}/2\pi$ & 7.158 GHz\\
 \hline
  Qubit decay rate & $1/T_1 = \Gamma_1$ & $1.11\ \mu$s$^{-1}$ \\
 \hline
  Qubit dephasing rate & $1/T_\varphi = \Gamma_\varphi$ & $1.32\ \mu$s$^{-1}$ \\
 \hline
 Resonator decay rate & $\kappa$ & $3.09\ \mu$s$^{-1}$ \\
 \hline
\end{tabularx}
    \caption{Device parameters.}
    \label{tab:DeviceParams}
\end{table}

In this work, we study the  polariton transitions as a function of varying coupler drive-strength. For each coupler power, we fix the frequency of the coupler tone at $\omega_d = \omega'_{ge,0}$, the dressed frequency of the qubit with zero photons in the resonator, and sweep the frequency of a weak probe tone to measure the transmission coefficient $|S_{21}|$. The probe frequency is swept across the ``mean frequency" $\omega'_{r,\text{mid}}=(\omega'_{r,e}+\omega'_{r,g})/2$, where $\omega'_{r,e(g)}$ is the singly dressed frequency of the resonator when the transmon is in the first-excited (ground) state. We vary the coupler powers from $-80$ dBm to $0$ dBm at the source ($\approx$ $-136\,$dBm to $-56\,$dBm at the device) (see Fig.~\ref{fig: experimental data - crossplot}). Following this, we also perform polariton spectroscopy as we vary the coupler frequency while keeping the coupler power fixed (see Fig.~\ref{fig:Spectrum_Multifreq}).

\section{\label{sec:Observations and results}Observations and Discussion} 
 \begin{figure}[!b]
  \centering
  \includegraphics[trim={0.0cm 0cm 0.3cm 0cm},clip,width=0.48\textwidth]{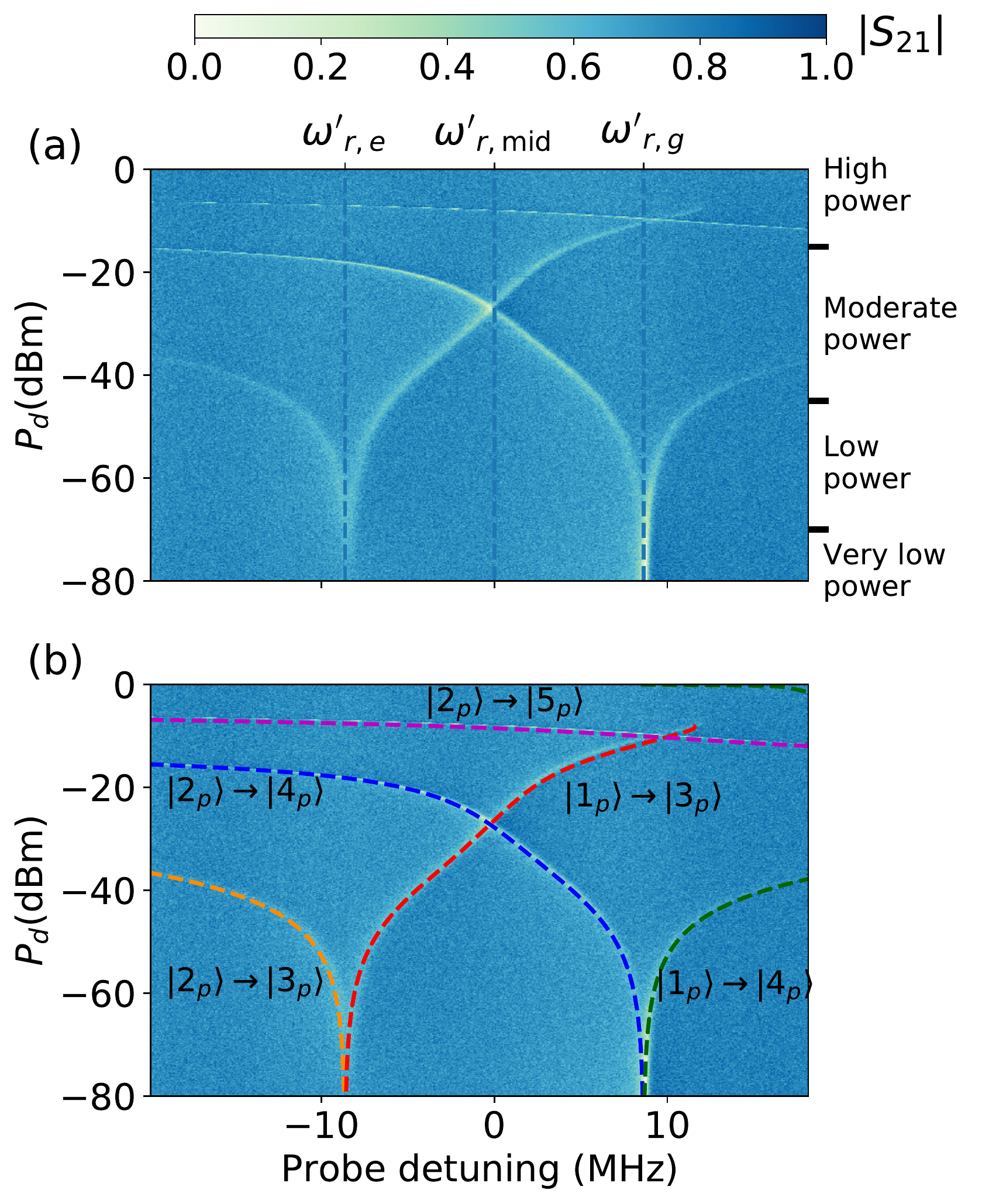}
  \caption{Polariton state spectroscopy with coupler drive power varied at $\omega_{{d}} = \omega_{ge,0}'$. (a) Experimental data with different regimes of coupler power marked (b) Experimental data with the dashed lines showing the simulated transition frequencies calculated by finding the eigenvalues of the Hamiltonian given in Eq.~(\ref{eq of Ham without probe}) using the parameters given in Table(\ref{tab:DeviceParams}).}
  \label{fig: experimental data - crossplot}
\end{figure}

\begin{figure}[tb]
  \centering
  \includegraphics[trim={0.2cm 0.3cm 0cm 0cm},clip,width=0.4\textwidth ]{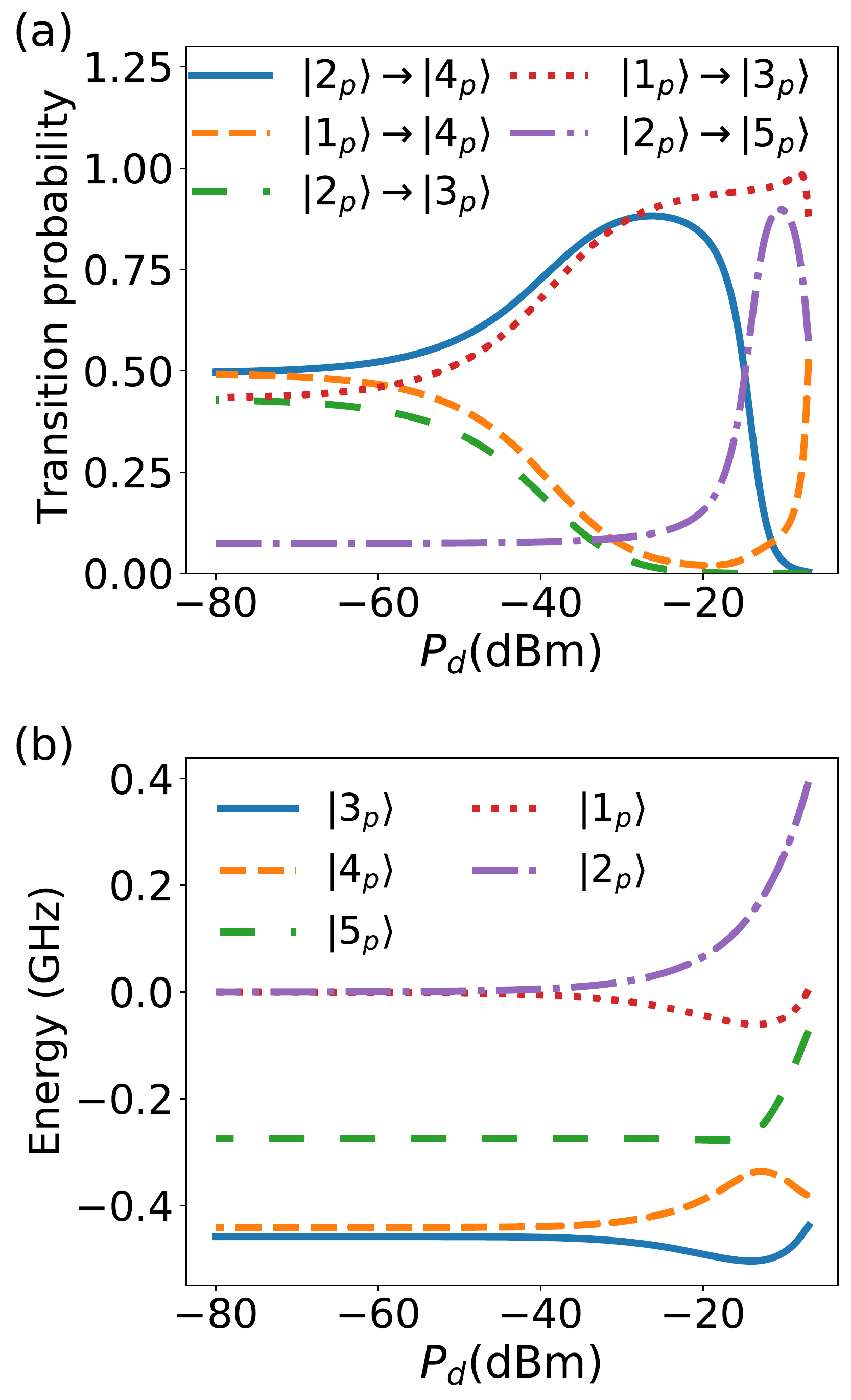}
  \caption{(a) Shows transition probability of all five visible transitions due to probe drive (b) energy of polariton states varying with the power of coupler drive in the dressed frame calculated by finding the eigenvalues of the Hamiltonian given in equation \ref{eq of Ham without probe} using the parameters given in Table(\ref{tab:DeviceParams}).}
  \label{contribution with drive}
\end{figure}


 Figure~\ref{fig: experimental data - crossplot}(a) shows the spectroscopy of polariton transitions for a fixed coupler frequency of $\omega_{{d}} = \omega_{ge,0}'$, as the coupler power $P_d$ is varied. We plot the quantity $\abs{S_{21}}$, which is the transmission amplitude normalized to the maximum observed value. The observed data can be classified into four regimes of coupler power, \textit{viz.}, very low power ($<-70$ dBm), low power ($-70$ dBm to $-45$ dBm), moderate power ($-45$ dBm to $-15$ dBm) and high power ($>-15$ dBm) regimes. 
 
 In Fig.~\ref{fig: experimental data - crossplot}(b), we overlay the measured data with results from an eigenmode calculation, which enables us to identify the various polaritonic transitions. Furthermore, we are able to qualitatively predict the intensity and linewidths of the observed lines using a full master equation simulation. 
The details of the eigenmode calculation and the master equation simulations are presented in Appendix \ref{Appendix_EigenmodeSimulation} and \ref{Appendix_MasterEquation}, respectively. In the following, we qualitatively explain the features observed in the experiment as shown in Fig.\ref{fig: experimental data - crossplot}, using the energies of the polariton states and the transition probabilities between these states, which are plotted as a function of coupler power in Fig.~\ref{contribution with drive}.

\subsection{\label{subsec:vlp}Very low power regime}

In the very low power regime, one observes only one line, which occurs at at $\omega'_{r,g}$, i.e. the dressed resonator frequency with transmon in the ground state. Here, only the state $\overline{\ket{g,0}}$ has a significant steady-state occupation, and the observed line corresponds to the $\overline{\ket{g,0}}\rightarrow\overline{\ket{g,1}}$ transition induced by the probe. In terms of polariton states, this corresponds to $\ket{1_p}\rightarrow\ket{4_p}$ and $\ket{2_p}\rightarrow\ket{4_p}$ transitions which are almost degenerate in this regime.

\subsection{\label{subsec:lp}Low power regime}

As the coupler power is increased, we move into the low power regime, where a second line appears. This line occurs at $\omega'_{r,e}$, \textit{i.e.} the dressed resonator frequency with transmon in the first-excited state, indicating that as the drive power increases, the state $\overline{\ket{e,0}}$ begins to get populated. To explain the two lines in terms of polariton states, we note that the non-dispersive nature of the higher transmon levels  enhances the dispersive shift  $\chi \approx (\omega_{ge,0}' - \omega_{ge,1}')/2\approx 2\pi \times 8.5$ MHz. The system in this regime can be analyzed within the two-level dispersive approximation, but using the  measured value of $\chi$ in Eq.~(\ref{disp_transition_freq}). For low drive powers such that $\Omega_d^2/\chi^2 \ll 1$, the transition frequencies are approximately given by
\begin{align}
    \omega_{13,p} &= \Tilde{\omega}_r - \chi + \Omega_d,\nonumber\\
    \omega_{14,p} &= \Tilde{\omega}_r + \chi+ \Omega_d,\nonumber\\
    \omega_{23,p} &= \Tilde{\omega}_r - \chi - \Omega_d,\nonumber\\
    \omega_{24,p} &= \Tilde{\omega}_r + \chi- \Omega_d.
    \label{eqn:polaritondisp}
\end{align}
Eq.~(\ref{eqn:polaritondisp}) explains why only two lines are initially visible in the low-power regime. To resolve any two transitions, the frequency difference between them should be greater than the resonator linewidth $\kappa/2\pi\approx  491$ kHz. The two spectral lines observed correspond to the nearly degenerate, and hence unresolved pairs of transitions $\ket{1_p}\rightarrow\ket{3_p} , \ket{2_p}\rightarrow\ket{3_p}$ and $\ket{1_p}\rightarrow\ket{4_p},\ket{2_p}\rightarrow\ket{4_p}$, which only differ by $\sim \Omega_d \lesssim \kappa$. This can also be seen in Fig~\ref{contribution with drive}(b) where the states $\ket{1_p}$ and $\ket{2_p}$, and hence the above-mentioned transitions, are nearly degenerate. 

We also observe that  the line around $\omega'_{r,g}$ is brighter than the one around $\omega'_{r,e}$. This can be explained by noting that, at low drive power and in the presence of transmon and resonator dissipation, the steady state population is predominantly in the ground state $\overline{\ket{g,0}}$.

With a further increase in power, the degeneracy of the $\ket{1_p}$ and $\ket{2_p}$ states, and hence of the two pairs of transitions, is lifted (see Fig.~\ref{contribution with drive}), and we observe four distinct lines. 


\subsection{\label{subsec:mp}Moderate power regime}

As one starts to move into the moderate power regime, we observe that the intensities of the $\ket{1_p}\rightarrow\ket{4_p}$ and $\ket{2_p}\rightarrow\ket{3_p}$ lines start to diminish. This agrees with the behavior of the corresponding transition probabilities, $\abs{\mel{\alpha}{a}{\beta}}^2$, which decrease with an increase in the coupler power, as seen in Fig.~\ref{contribution with drive}(a). Therefore only the $\ket{1_p}\rightarrow\ket{3_p}$ and $\ket{2_p}\rightarrow\ket{4_p}$ transitions are visible as the drive power increases.

In this regime, the dispersive two-level approximation breaks down for our system. Equations~(\ref{disp_transition_freq}) predict that the $\ket{1_p}\rightarrow\ket{3_p}$ and $\ket{2_p}\rightarrow\ket{4_p}$ transition frequencies only asymptotically converge as the drive power is increased. Contrary to this prediction, we observe a distinct crossing of the two lines around a drive power of $-26$ dBm. The presence of a distinct crossing can be qualitatively explained by the unusually large value of $\chi$ in our device, which is made possible by the non-dispersive coupling of the higher transmon levels to the resonator. We discuss this in greater detail in Sec. \ref{sec:Comparison with dispersive case}.

\subsection{\label{subsec:hp}High power regime}

As one increases the power of the coupler drive even further, going into the high-power regime, a fifth line is observed (see Fig.~\ref{fig: experimental data - crossplot}).
This fifth line can be explained using Fig.~\ref{contribution with drive}(a), where the transition probability between $\ket{2_p}$ and a fifth polariton state $\ket{5_p}$ becomes appreciable at high drive powers. Furthermore, the frequency of this transition falls within the probed range of frequencies only at high powers. The crossing between $\ket{2_p}\rightarrow\ket{5_p}$ and $\ket{1_p}\rightarrow\ket{3_p}$ transitions  around $-10$ dBm is also predicted by numerical calculations of the polariton state energies as seen in Fig.~\ref{fig: experimental data - crossplot}(b).

\subsection{\label{subsec:specdifffreq}Spectroscopy at different coupler frequencies}

\begin{figure}[!htb]
  \centering
  \includegraphics[trim={0cm 0cm 0cm 0cm},clip,width=0.48\textwidth ]{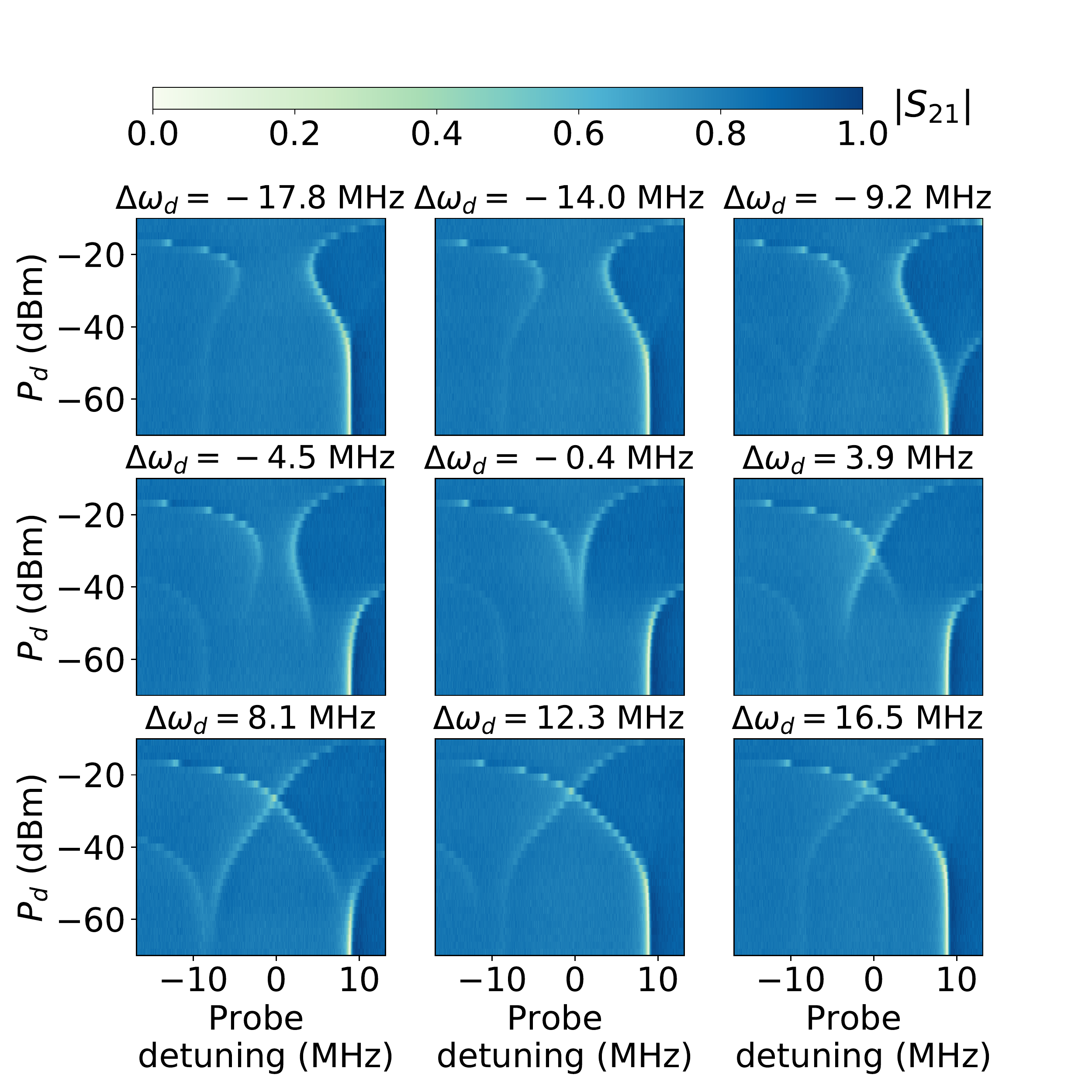}
  \caption{Spectroscopy plot at different frequencies of coupler drive detunings}
  \label{fig:Spectrum_Multifreq}
\end{figure}

Previously, we discussed spectroscopic measurements when the coupler drive frequency $\omega_{d}$ was resonant with the $\overline{\ket{g,0}} \rightarrow \overline{\ket{e,0}}$ transition. Here, we consider polariton spectra measured with different values of $\Delta\omega_{{d}} = \omega_{d} - \omega_{ge,\text{mid}}'$ where $(\omega_{ge,\text{mid}}' = \omega_{ge,0}' + \omega_{ge,1}')/2$, as shown in Fig.~\ref{fig:Spectrum_Multifreq}. From these plots, we observe that the crossing between the $\ket{1_p}\rightarrow\ket{3_p}$ and $\ket{2_p}\rightarrow\ket{4_p}$ lines can be observed only for $\omega_{{d}} \gtrsim (\omega_{ge,0}' + \omega_{ge,1}')/2$. This observation is consistent with our numerical simulations (see Appendix~\ref{Appendix_MasterEquation}).

\begin{figure}[!htb]
  \centering
  \includegraphics[width=0.48\textwidth]{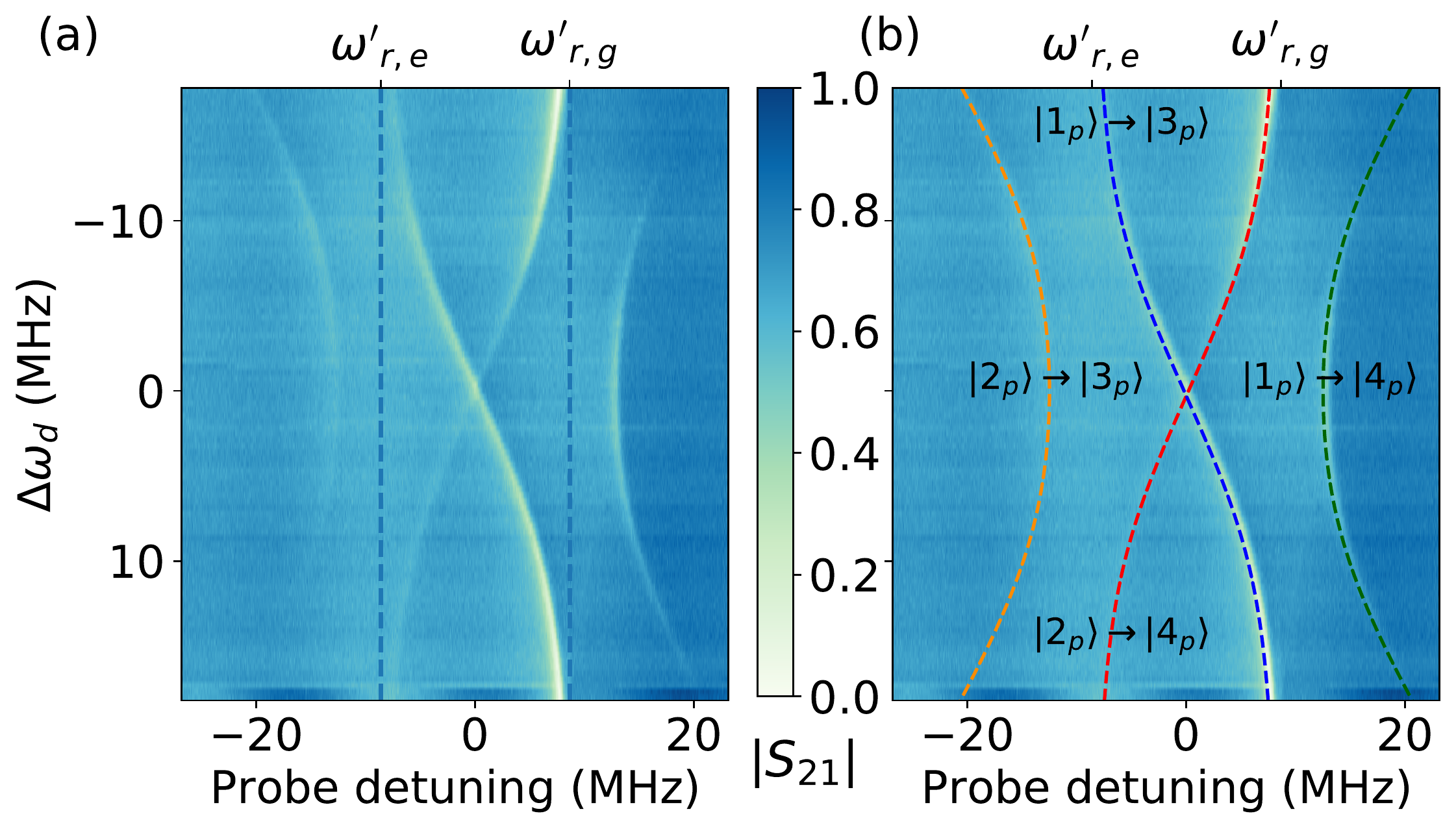}
  \caption{Polariton state spectroscopy with $\omega_{{d}}$ varied at coupler drive power of -40 dBm (a) Experimental data (b) Experimental data with eigenmode simulation calculation similar to Fig.~(\ref{fig: experimental data - crossplot})}
  \label{fig: experimental data - xplot}
\end{figure}

In order to gain further insight into the effect of coupler tone frequency, we performed spectroscopy of the polariton states by sweeping the coupler frequency while keeping the coupler power fixed. The measured data for a coupler power of $-40$ dBm is shown in Fig.~\ref{fig: experimental data - xplot}(a) and a comparison with the eigenmode calculation is shown in Fig.~\ref{fig: experimental data - xplot}(b). We observe that the spectrum at a coupler tone frequency $\omega_{{d}} = \omega_{ge,\text{mid}}'$ resembles a Mollow triplet. This value of $\omega_{{d}}$ corresponds to the point where $\ket{1_p}\rightarrow\ket{3_p}$ and $\ket{2_p}\rightarrow\ket{4_p}$ transitions become degenerate. Moreover, the triplet is observed at this specific value of $\omega_{{d}}$ for a broad range of drive powers. At coupler frequencies away from this point, we observe four lines in the spectrum corresponding to the four polaritonic transitions.

\section{\label{sec:Comparison with dispersive case}Comparison with dispersive case} 
In this section, we discuss the role of the non-dispersive coupling of the higher transmon levels in the observed spectrosopic data.

\subsection{\label{subsec:meqsim}Master equation simulations}

First, we qualitatively compare master equation simulations of our device in the non-dispersive regime, with a typical transmon-resonator system in the dispersive regime, where the bare transmon frequency is $1$ GHz below the resonator, as shown in Fig.~\ref{disp_vs_nondisp}. All other parameters are the same  for both systems and are listed in Table~\ref{tab:DeviceParams}. The details of the master equation and the numerical methods used are given in Appendix~\ref{Appendix_MasterEquation}.

\begin{figure}[!htb]
  \centering
   \includegraphics[width=0.9\columnwidth ]{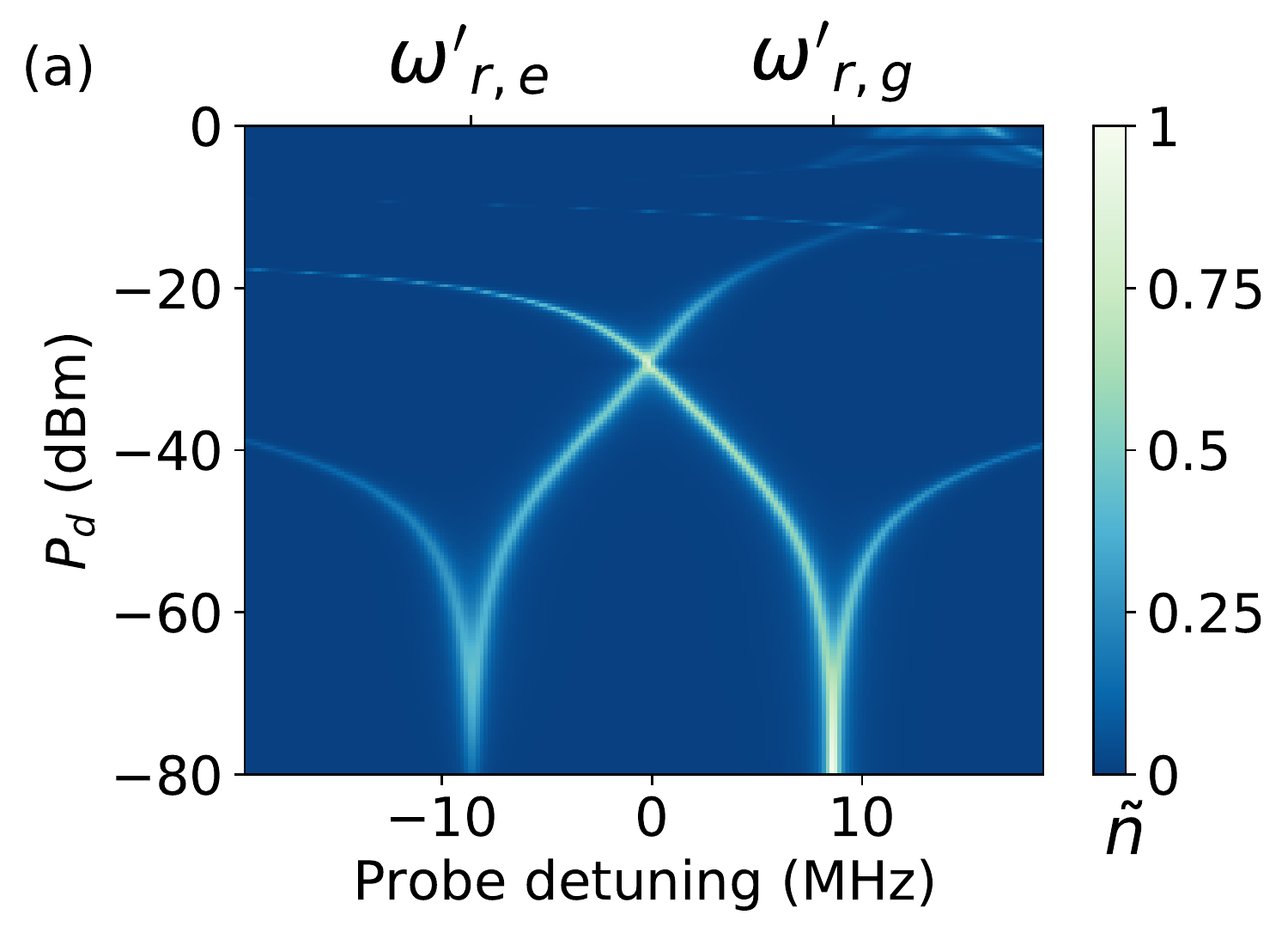}
     \includegraphics[width=0.9\columnwidth]{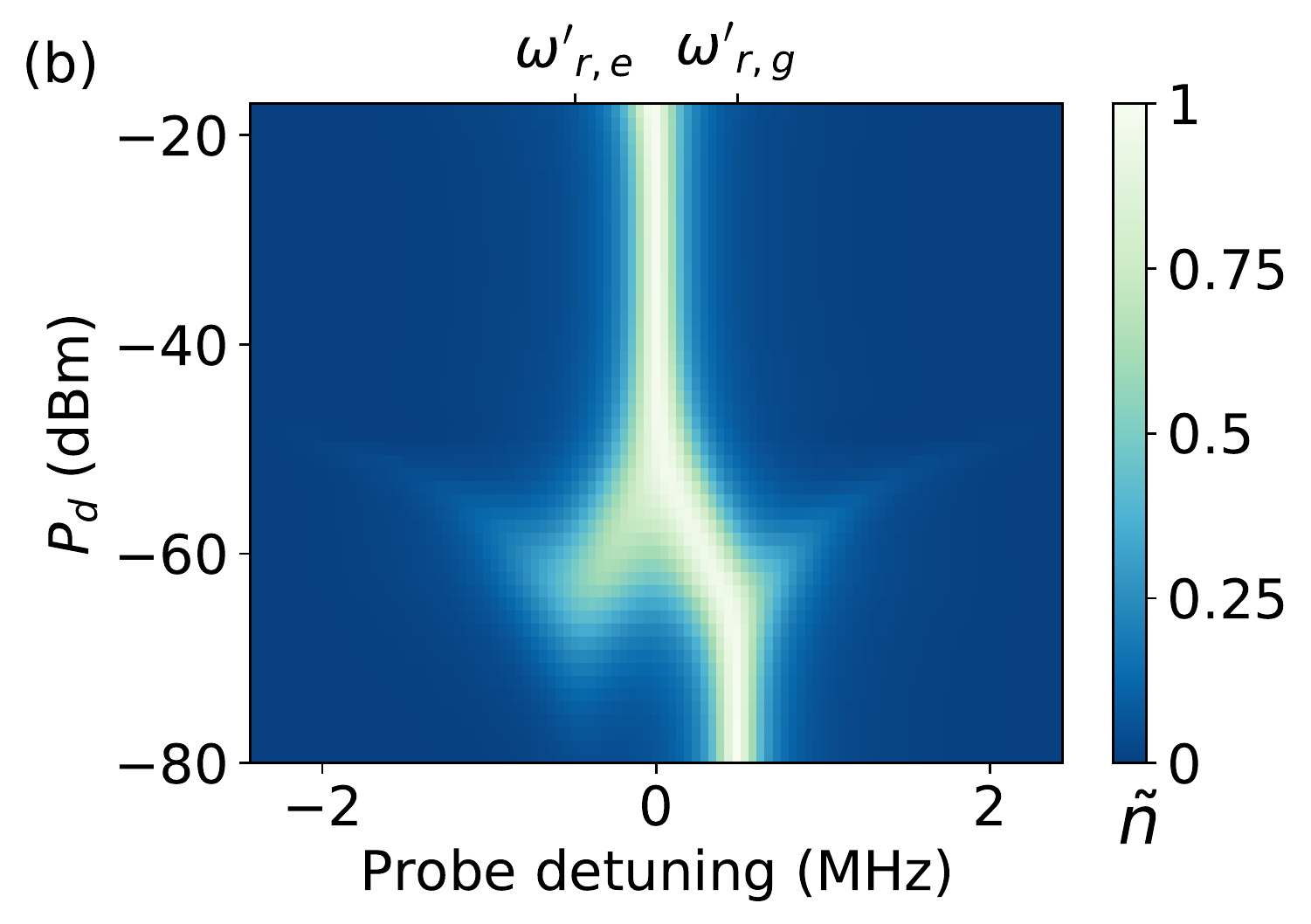}
  \caption{Master Equation simulation of spectroscopy of the system with (a) Non-Dispersive Coupling, (b) Dispersive Coupling.}
  \label{fig:SpecDispCoup}
\end{figure}

The simulated spectra for the two systems are shown in Fig.~\ref{fig:SpecDispCoup}(a,b). For the simulations, we plot the normalized mean number of photons in the resonator in steady state, defined as 
\begin{eqnarray}
    \tilde{n} = \frac{\ev{\hat{a}^\dagger \hat{a}}-\ev{\hat{a}^\dagger \hat{a}}_\mathrm{min}}{\ev{\hat{a}^\dagger \hat{a}}_\mathrm{max}-\ev{\hat{a}^\dagger \hat{a}}_\mathrm{min}},
    \label{eqn:ntilde}
\end{eqnarray}
where $\ev{\hat{a}^\dagger \hat{a}}$ is the mean resonator occupation in steady state and the maximum and minimum values are taken over the range of parameters scanned. The location of the lines, their intensities, and linewidths in Fig.~\ref{fig:SpecDispCoup}(a) are consistent with the measured data shown in Fig.~\ref{fig: experimental data - crossplot}. A noticeable difference between the spectra in Fig.~\ref{fig:SpecDispCoup}(a) and (b) is the absence of a distinct crossing of the lines in the latter case. Instead, we observe the two lines approaching each other in frequency, becoming indistinguishable beyond $-55$ dBm of coupler power.

\subsection{\label{subsec:qualexp}Qualitative explanation for distinct crossing}

The observation of a distinct crossing in the experiment is the result of a large value of $\chi$ which, in our device, originates from the non-dispersive coupling of the $\ket{e}\rightarrow\ket{f}$ transition to the resonator. Despite the non-dispersive origin of this large $\chi$, we can qualitatively predict a crossing through analytical calculations in  a multilevel dispersive approximation, while using a large value of $\chi$. To do so, we extend the work of Ref.~\cite{Gu2016} to include the effect of higher levels along the lines of Ref.~\cite{koch2007charge}. The central idea is that, for the data shown in Fig.~\ref{fig: experimental data - crossplot}, the coupler drive frequency is near-resonant with $\overline{\ket{g,n}}\rightarrow\overline{\ket{e,n}}$ transitions but off-resonant from $\overline{\ket{e,n}}\rightarrow\overline{\ket{f,n}}$ transitions. This is because of the fact that, for moderate drive powers, we have $\Omega_d\ll \alpha$ where $\alpha = \omega_{e,f} - \omega_{g,e}$ is the anharmonicity of the transmon. As a result, the off-resonant driving of higher transmon transitions by the coupler can be treated using perturbation theory. The details of this calculation are presented in Appendix~\ref{Appendix:Derivation_SOPT}. Using this approach, we find that the frequencies of the $\ket{1_p}\rightarrow\ket{3_p}$ and $\ket{2_p}\rightarrow\ket{4_p}$ polaritonic transitions are given by 
\begin{align}
    \label{eqn:transition_freqs}
    \omega_{13,p} = \Tilde{\omega}_r - (\sqrt{\chi^2 + \Omega_d^2} - \Omega_d) + \frac{\Omega_d^2 \cos\theta_1}{\alpha},\nonumber\\
    \omega_{24,p} = \Tilde{\omega}_r  + (\sqrt{\chi^2 + \Omega_d^2} - \Omega_d) - \frac{\Omega_d^2 \cos\theta_1}{\alpha},
\end{align}
where $\tan\theta_1=-\Omega_d/\chi$. The two frequencies become degenerate
($\omega_{13,p}=\omega_{24,p}$) when
\begin{equation}\label{eq:SOPT_Crossing}
    \sin\theta_1 + \frac{\Omega_d \sin(2\theta_1)}{2\alpha} = 1.
\end{equation}
This equation reduces to the case of a transmon truncated to a two-level system in the limit that $\alpha \rightarrow \infty$. In this case, the crossing condition becomes $\theta_1=\pi/2$, which is only satisfied as $\Omega_d \rightarrow \infty$ and hence the crossing is not observed.

For a multi-level system like a transmon, where $\alpha$ is finite, we can obtain an approximate crossing condition from Eq.~(\ref{eq:SOPT_Crossing}). For $\Omega_d^2\gg \chi^2$, we can substitute $\sin\theta_1\approx 1-\chi^2/(2\Omega_d^2)$, $\cos\theta_1\approx -\chi/\Omega_d$ in Eq.~(\ref{eq:SOPT_Crossing}) and obtain the crossing condition $\Omega_d^2\approx -\chi\alpha/2$. In addition, in order to observe a distinct crossing, we require that the crossing condition must be satisfied for drive powers such that the two transitions are well resolved. Since the resolution of the observed transitions is limited by the resonator linewidth $\kappa$, the two transitions are resolvable when their frequency difference close to the crossing is greater than $\kappa$, i.e. $\sqrt{\chi^2 + \Omega_d^2} - \Omega_d > \kappa$. By  substituting the crossing condition $\Omega_d^2\approx -\chi\alpha/2$ into this inequality, we arrive at the condition $\abs{\chi} \gtrsim (\kappa \sqrt{2\abs{\alpha}})^{2/3}$ that satisfies both the above requirements.

For a transmon-resonator system with typical values of $\alpha/2\pi \sim -300$ MHz and $\kappa/2\pi \sim 0.5\text{ -- }10$ MHz, this leads to a required value of $\chi/2\pi \gtrsim 5.3\text{ -- }39$ MHz. In our device, the non-dispersive coupling of the upper transmon levels to the resonator leads to $\chi/2\pi \approx 8.5$ MHz, which for a $\kappa/2\pi \approx 0.5$ MHz, enables us to observe a distinct crossing. On the other hand, for a transmon system coupled in the dispersive regime, usually $\chi/2\pi \lesssim 2$ MHz, making the observation of a distinct crossing difficult. 

\begin{figure}[!htb]
  \centering
     \includegraphics[width=0.9\columnwidth]{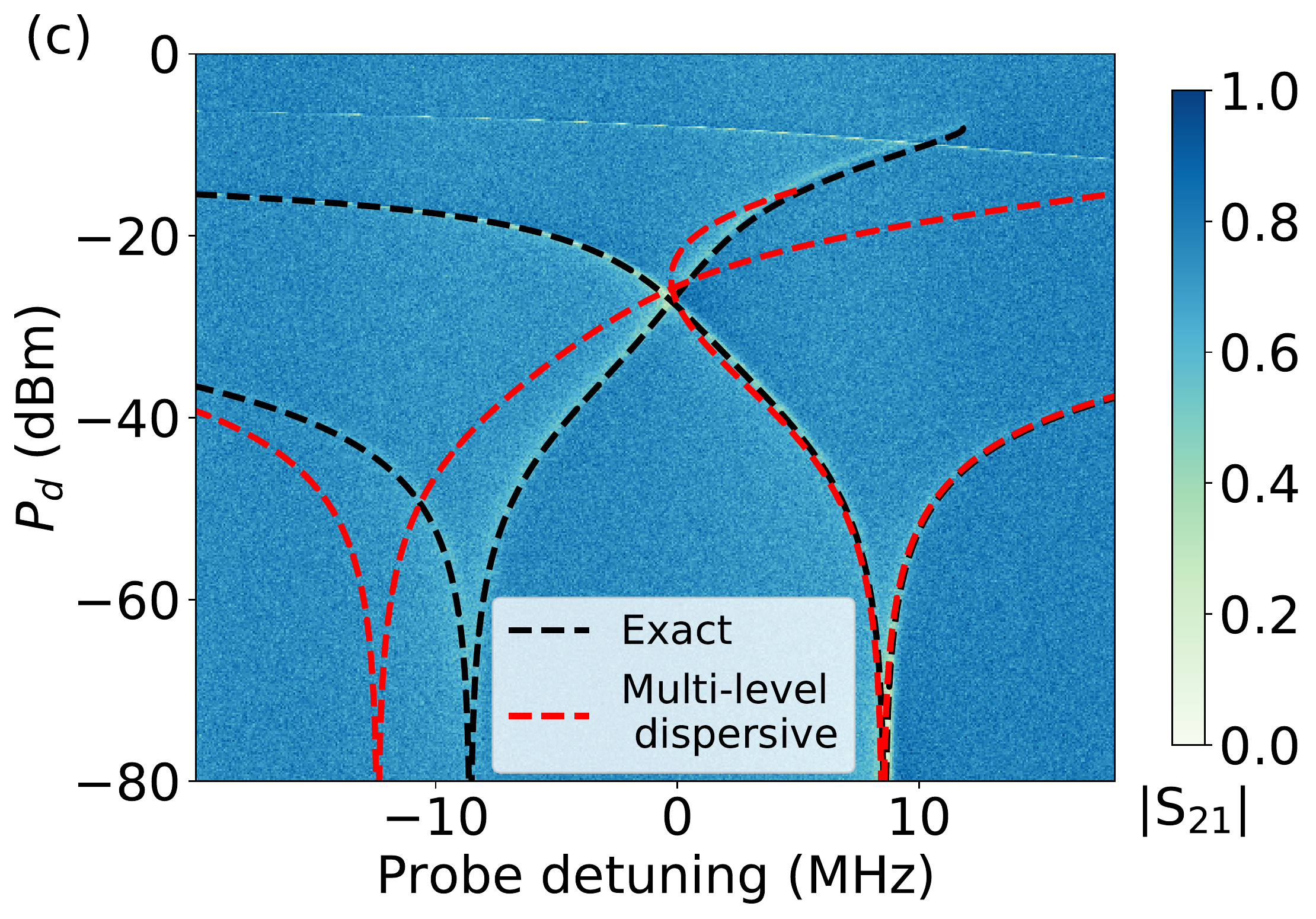}
  \caption{Comparison of eigenmode simulation for the  Hamiltonian vs Multilevel dispersive Hamiltonian in Eq.~(\ref{Appendix:multilevel dispersive}).}
  \label{fig:comp_mdisp_exact}
\end{figure}

We note that a similar spectrum and crossing, as shown in Fig.~\ref{fig: experimental data - crossplot}, was also reported in Ref.~\cite{PhysRevLett.124.070401}. However, in that work, the device operated in a fully dispersive regime, and hence the observed spectra and crossing could be fully explained within the dispersive approximation. In contrast, in our system, although we are able to qualitatively explain the existence of a crossing using the multilevel dispersive approximation, we find that it fails to achieve a good quantitative agreement in master equation simulations as shown in Fig.~\ref{fig:comp_mdisp_exact}. Instead, as shown in Fig.~\ref{fig:SpecDispCoup}(a), we obtain excellent agreement when using the full non-dispersive JC interaction in our simulations. The different operating regimes of the device in Ref.~\cite{PhysRevLett.124.070401} and in our work is illustrated in Fig.~\ref{disp_vs_nondisp}(a).

\section{\label{sec:conclusion}Conclusion}

We theoretically and experimentally studied the properties of polariton states in a transmon-resonator system operating in a non-dispersive regime. The non-dispersive coupling arises from the fact that the frequency of $\ket{e}\rightarrow\ket{f}$ transition of the transmon lies close to the resonator frequency, resulting in a value of $g_1/\Delta_1\approx 0.47$. By introducing a coupler drive on the transmon, we generated polariton states.  We spectroscopically studied the  polarition transitions using a weak probe field on the resonator. By using eigenmode analysis, and master equation simulations, we were able to explain the origin of the observed spectral lines and their intensities. We found that the observed lines differed significantly from those expected in the dispersive regime. In particular, at moderate coupler drive powers, we observed a distinct crossing between two lines. Using perturbation theory, we derived a condition on $\chi$ required to observe a crossing, and  showed that the large value of $\chi/2\pi\approx8.5$ MHz in our device satisfied this condition. We also showed that this condition was difficult to be satisfied in the usual dispersive regime.

\begin{acknowledgements}
The authors acknowledge the support of Ministry of Electronics and Information Technology, Government of India, under the Centre for Excellence in Quantum Technology grant to Indian Institute of Science. The authors also acknowledge support from the Department of Science and Technology, India, via the QuEST program. The authors acknowledge the  fabrication and characterisation facilities NNFC and MNCF at CeNSE, IISc, as well as fabrication facilities supported by the Institute of Eminence grant to IISc by Ministry of Education, Govt. of India. AM acknowledges the support of Ministry of Education, Government of India. SH acknowledges the support of Kishore Vaigyanik Protsahan Yojana, Department of Science and Technology, Government of India. AS acknowledges the support of a CV Raman Post-Doctoral Fellowship, IISc. 
\end{acknowledgements}

\appendix

\section{\label{sec:Appendix_DeviceFabrication}Device Fabrication}

The device was fabricated in three steps of lithography.  
The first step involved patterning of the gold alignment markers. A fresh 2-inch Si wafer was coated with an optical resist S1805, spun at 4000 rpm for 45 sec and  1 minute of baking at 110$^{\circ}$C. Then the alignment marks were written using Heidelberg direct-write laser writer followed by development using AZ786 developer. Post development, the wafer went through the process of $\mathrm{O}_2$ plasma ashing to remove the extra photoresist. A thin film of gold was deposited using an e-beam evaporator (Leybold) followed by a liftoff process using acetone.  A second step of photolithography was used to pattern the CPW transmission line, resonator and ground plane in Aluminum. The deposition of Al was done in an e-beam evaporator followed by a liftoff process using acetone. The final step of patterning the Josephson junctions involved electron beam lithography in a 30kV Raith eLINE tool. The samples were first coated with PMMA 950 C6 resist and baked for 15 minutes at 180$^{\circ}$c. Following e-beam lithography, and development using MIBK: IPA(1:3) for 40 sec and IPA for 1 min, the Josephson junctions were formed using a bridgeless junction technique \cite{potts2001cmos}. The intermediate oxidation step between the two Al depositions was done at 700 mTorr pressure for 20 min. 
\section{\label{sec: Time domain experiment setup}Setup for time domain measurements} 
\begin{figure}[h!tb]
  \centering
  \includegraphics[trim={0cm 0cm 0cm 0cm},clip,width=0.45\textwidth]{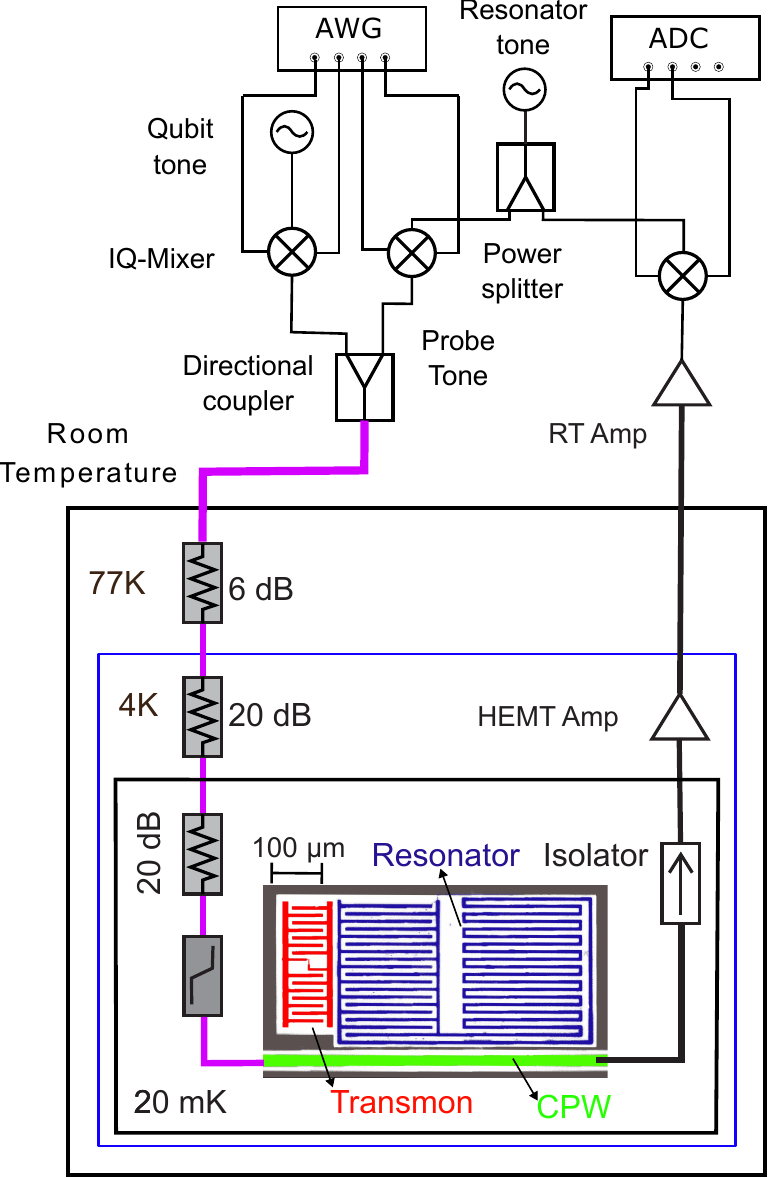}
  \caption{Schematic diagram for the time domain measurement. The time domain pulses were generated using an AWG (arbitrary waveform generator) for both the transmon and the resonator. The output signals were down-converted and digitized using an ADC (Analog-to-digital converter) with a $1\,$GSa/s sampling rate.}
  \label{Schematics time domain}
\end{figure}

We measured the relaxation and dephasing times of the transmon, $\text{T}_1$ and $\text{T}_{\phi}$ respectively, using time domain measurements. The setup used to carry out the time domain measurements is shown in Fig.~(\ref{Schematics time domain}). We followed standard time domain measurement techniques as described in Ref~\cite{reed2010high, suri2015transmon} using a high-power readout scheme. 

\section{\label{Appendix_EigenmodeSimulation}Eigenmode analysis}

We recall that the Hamiltonian in the frame of the coupler drive is given by 
\begin{eqnarray}\label{app_eq of Ham without probe}
   \frac{\hat{H}_\text{rot}}{\hbar} = &\delta_r \hat{a}^{\dagger}\hat{a}  + \sum_{n}\delta_n\ket{n}\bra{n} +  g_0(\hat{a}^{\dagger}\hat{b} + \hat{a}\hat{b}^\dagger)\nonumber\\
   &+ \Omega_d (\hat{b} + \hat{b}^{\dagger}),
\end{eqnarray}
where the symbols are defined in Section~\ref{sec:Theoretical model}. To find the energy of polariton states, we diagonalise Hamiltonian~\ref{app_eq of Ham without probe}. We note that the transition frequencies in the lab frame can be obtained from the eigenfrequencies of this Hamiltonian upon shifting them by the coupler frequency. This point becomes clear when computing the transition matrix elements in the presence of a probe, which is discussed below.

In the additional presence of a probe, the full Hamiltonian $H_p$ of the system can be written in the frame of the coupler drive as 
\begin{equation}
    \hat{H}_p = \hat{H}_\text{rot} + \hbar \Omega_p (\hat{a} e^{i\delta_p t} + \ad e^{-i\delta_p t}),
\end{equation}
where $\Omega_p$ is the probe strength, $\omega_p$ is the probe frequency, and $\delta_p = \omega_p - \omega_d$. 

As this probe is considered to be weak, we can neglect its effect on the eigenspectrum of the Hamiltonian. However, we do not neglect it in the context of transitions between eigenstates of $H_\text{rot}$, for example $\ket{\alpha}, \ket{\beta}$ with respective energies $\hbar \omega_\alpha, \hbar \omega_\beta$. The corresponding transition matrix element is given by
\begin{align}
    \tau &= \mel{\alpha(t)}{(H_p/\hbar)}{\beta(t)}\nonumber\\
        &= \Omega_p \left(\bra{\alpha}e^{i\omega_\alpha t}\right){(\hat{a} e^{i\delta_p t} + \ad e^{-i\delta_p t})}\left(e^{-i\omega_\beta t}\ket{\beta}\right) \nonumber \\
        &=  \Omega_p (e^{-i(\omega_\tau-\delta_p) t} \mel{\alpha}{\hat{a}}{\beta} + e^{i(\omega_\tau-\delta_p) t} \mel{\alpha}{\ad}{\beta}),
\end{align}
where $\hbar \omega_\tau = \hbar \left(\omega_\beta-\omega_\alpha\right)$ is the  difference between the energies of the states in the coupler drive frame. From this, we see that to obtain a non-zero time-averaged value of $\tau$, the probe frequency must equal the transition frequency in the lab frame, which is given by $\omega_d+\omega_\tau$, which is the condition for energy conservation.

For all the simulations we used the \verb+QuTiP+~\cite{johansson2012qutip} package in python, and considered 4 levels in both transmon and the resonator. 
The Rabi frequency $\Omega_d$ is related to the amplitude of the drive as 
$\Omega_d = \beta V$ where $\beta$ is the coupling parameter between the drive and resonator. The voltage $V$ can be converted to the power applied to the device using the relation 
$V = \sqrt{Z\times 10^{(P_c-att-30)/10}}$ where $Z$ is the impedance of the transmission line (CPW), and $P_c$ is the drive's power at the source in dBm and $att$ is the attenuation present in the line. 
Using the above relation, all the unknown parameters can be absorbed in one constant $C$ such that $\Omega_d = C 10^{P_c/20}$ where $C=\beta \sqrt{Z\times 10^{-(att+30)/10}}$. The parameter $C$ was tuned along with $\omega_{ge}$, $g$ and $\alpha$ to get the agreement between the experimental data and with simulation results, and the value of $C$ was found to be 0.562.

\section{\label{Appendix_MasterEquation}Details of master equation simulations}

While eigenmode analysis enables us to identify the observed transitions and qualitatively predict their intensities and linewidths, in order to get a quantitative prediction, we use master equation simulations. Starting from the lab-frame Hamiltonian including both the coupler and the probe fields, we transform to an interaction picture through the unitary operator 
\begin{equation}
  \hat{U}'_{\mathrm{rot}} = e^{i\omega_p t (\hat{a}^{\dagger}\hat{a}) + i\omega_d t ( \hat{b}^{\dagger}\hat{b})},
  \label{app_unitary probe drive}
\end{equation}
resulting in an interaction Hamiltonian
\begin{multline}
\label{app_eqn:hrot}
   \frac{\hat{H}'_\text{rot}}{\hbar} = \delta_r \hat{a}^{\dagger}\hat{a}  + \sum_{n}\delta_n\ket{n}\bra{n} + 
   \Omega_d (\hat{b} + \hat{b}^{\dagger}) +
   \Omega_p (\hat{a} + \hat{a}^{\dagger})\\
   + g_0(e^{i\Delta t}\hat{a}^{\dagger}\hat{b} + e^{-i\Delta t}\hat{b}^{\dagger}\hat{a}),
\end{multline}
where $\Delta = \omega_p - \omega_d$. 

To include the environmental effects like the decay of transmon and resonator, the master equation simulations were performed including Lindblad terms describing the decay (rate $\Gamma_1$) and dephasing ($\Gamma_\phi$) of the transmon, and the decay of the resonator ($\kappa$). The complete master equation is 
\begin{equation}
    \dot{\hat{\rho}}=-\frac{i}{\hbar}[\hat{H}'_\text{rot},\hat{\rho}] + \kappa \mathcal{D}[\hat{a}]\hat{\rho} + \Gamma_1\mathcal{D}[\hat{b}]\hat{\rho} + \Gamma_{\phi}\mathcal{D}[b^{\dagger}b]\hat{\rho},
\end{equation}
where $\mathcal{D}[\hat{A}]\hat{\rho} = \hat{A}\hat{\rho} \hat{A}^{\dagger} - \frac{1}{2}(\hat{A}^{\dagger}\hat{A}\hat{\rho} + \hat{\rho} \hat{A}^{\dagger}\hat{A})$ for an operator $\hat{A}$.

The simultaneous presence of both coupler and probe means that there is no rotating frame in which the Hamiltonian can be made time-independent. Since this equation is time-dependent, finding steady-state solutions to it can be computationally expensive. To decrease the computational resources required, we make use of the method of Matrix Continued Fractions \cite{risken_fokker-planck_1989, sze_meng_tan_quantum_nodate} as our time-dependent terms are sinusoidal with a single frequency. In this method, the density matrix is reshaped into a vector and the master equation is expressed as a vector differential equation as 
\begin{equation}
    \dot{\hat{\rho}} = (\Lv_0 + \Lv_{1}e^{i\Delta t} + \Lv_{-1}e^{-i\Delta t})\hat{\rho},
\end{equation}
where $\Lv_{(-1,0,1)}$ are Liouvillian superoperators represented as matrices acting on the vector $\hat{\rho}$. This equation can be solved iteratively by expanding $\hat{\rho}$ into its frequency components. We write $\hat{\rho}(t)$ as
\begin{equation}
	\hat{\rho}(t) = \sum_{n\in\Z} \hat{\rho}_n e^{i n \Delta t}.
\end{equation}
Substituting this ansatz into the master equation gives
\begin{equation}
	\sum_n i n \Delta \hat{\rho}_n e^{i n \Delta t} = \sum_n (\Lv_0 + \Lv_{1}e^{i\Delta t} + \Lv_{-1}e^{-i\Delta t})\hat{\rho}_n e^{i n \Delta t},
\end{equation}
through which a recursion relation of the form,
\begin{equation}
	\Lv_1 \hat{\rho}_{n-1} + (\Lv_0 - i n \Delta) \hat{\rho}_n + \Lv_{-1} \hat{\rho}_{n+1} = 0,
\end{equation} 
between different frequency components of $\hat{\rho}$ is obtained. This equation needs to be solved \textit{via} the method of Matrix Continued Fractions. First, for $n>0$, we set $\hat{\rho}_n = S_n \hat{\rho}_{n-1}$. Substituting and simplifying, we get
\begin{equation}
	S_n = -((\Lv_0 - i n \Delta) + \Lv_{-1} S_{n+1})^{-1} \Lv_{1},
\end{equation} 
through which we can obtain $S_1$ by setting a large $N$ such that $S_n = 0~ \forall~ n>N$. This method converges quickly as it is based in the regular continued fraction method that can be proven to converge ``exponentially'' quickly.

A similar treatment for $n<0$ with $\hat{\rho}_n = T_n \hat{\rho}_{n+1}$ gives
\begin{equation}
	T_n = -((\Lv_0 - i n \Delta) + \Lv_{1} T_{n-1})^{-1} \Lv_{-1},
\end{equation} 
from which $T_{-1}$ can be obtained by setting $T_{-N} = 0$.

Finally, we can substitute everything back into the recursion relation at $n=0$ to get
\begin{equation}
	(\Lv_{-1}S_1 + \Lv_0 + \Lv_1 T_{-1})\hat{\rho}_0 = 0,
\end{equation} 
which is solved easily to get $\hat{\rho}_\textrm{ss} = \ev{\hat{\rho}}_t = \hat{\rho}_0$, and hence $\hat{\rho}_n$.

A partial implementation of this method can be found in \verb+QuTiP+~\cite{johansson2012qutip} for $\Lv_1 = \Lv_{-1}$. For our simulations, we have extended the method to treat the case when $\Lv_1 \neq \Lv_{-1}$

\begin{figure}[tb]
  \centering
  \includegraphics[trim={0cm 0cm 0cm 0cm},clip,width=0.48\textwidth ]{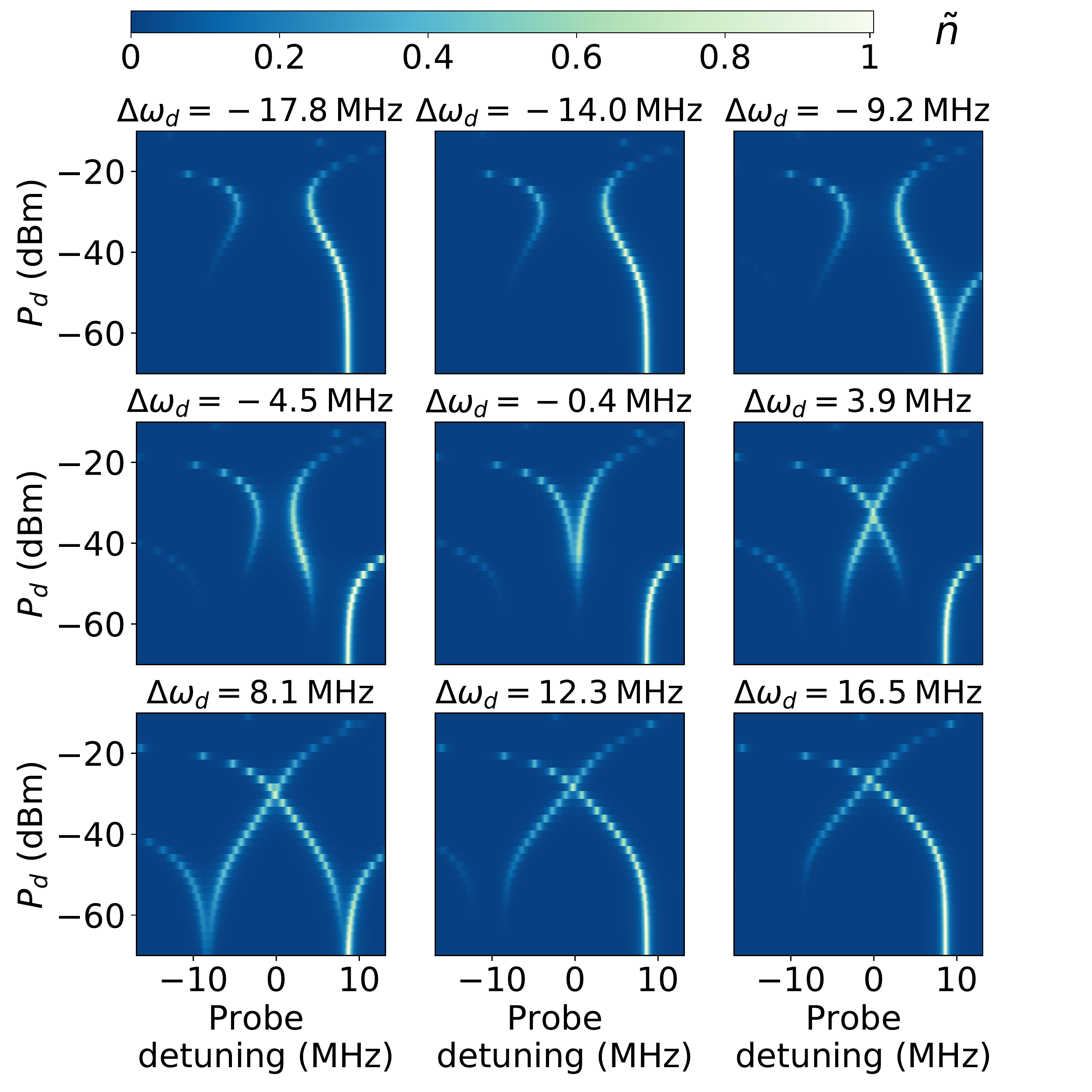}
  \caption{Master Equation simulations corresponding to spectroscopy plot Fig.~\ref{fig:Spectrum_Multifreq} at different coupler drive detunings, as described in Appendix~\ref{Appendix_MasterEquation}. The plot shows the normalized value for the occupation of photons in the resonaotor.}
  \label{fig:sim_Spectrum_Multifreq}
\end{figure}

For our system, we  write the terms in the master equation as
\begin{align}
    &\Lv_0 = -\frac{i}{\hbar}[\hat{H}'_{\text{rot},0},\cdot] + \kappa \mathcal{D}[\hat{a}]\cdot + \Gamma_1\mathcal{D}[\hat{b}]\cdot + \Gamma_{\phi}\mathcal{D}[\hat{b}^{\dagger}\hat{b}]\cdot,\nonumber\\
    &\Lv_1 = -\frac{i}{\hbar}[g_0\hat{a}^\dagger \hat{b},\cdot],\;
    \Lv_{-1} = -\frac{i}{\hbar}[g_0 \hat{a} \hat{b}^\dagger,\cdot],
\end{align}
where $\hat{H}'_{\text{rot},0}$ is the time-independent part of the Hamiltonian $\hat{H}'_{\text{rot}}$ in Eq.~(\ref{app_eqn:hrot}). Solving these equations as described above, we find the time-independent steady-state component $\rho_\mathrm{ss}$ and the associated mean number of photons in the resonator mode $\ev{\ad \hat{a}}=\mathrm{Tr}[\ad \hat{a}\, \hat{\rho}_\mathrm{ss}]$, using which we compute $\tilde{n}$ according to Eq.~(\ref{eqn:ntilde}). In the experiment, we can measure the coefficient of power transmission, which is proportional to the steady-state resonator occupation. Hence, we plot the latter observable as a proxy for the spectrum in our numerical simulations.

In Fig.~\ref{fig:sim_Spectrum_Multifreq}, we show simulated spectra for the parameters of the experimental data presented in Fig.~\ref{fig:Spectrum_Multifreq}. The simulations reproduce the observed spectra very well and provide further confirmation of the accuracy of our modelling.

\section{\label{Appendix:Derivation_SOPT} Perturbation theory calculations to account for higher levels in the transmon}
In this appendix, we show the detailed calculations that help us explain the existence of a crossing in a multilevel non-dispersive system. To do this, we extend the work of Ref.~\cite{Gu2016} to consider the effect of higher levels in the transmon. Applying a multilevel dispersive rotation to the Hamiltonian~(Eq.~\ref{eq of Ham without probe}), we obtain the Hamiltonian written in the singly dressed basis as~\cite{koch2007charge}
\begin{multline}
    \frac{\hat{H}_\text{disp,rot}}{\hbar} \approx \sum_n \delta'_n \overline{\op{n}{n}} + \delta'_r \ad \hat{a} - \chi_{01}\ad \hat{a} \overline{\op{0}{0}} \\
    + \sum_{n>0} (\chi_{n-1,n} - \chi_{n, n+1})\ad \hat{a} \overline{\op{n}{n}} + \Omega_d (\hat{b} + \hat{b}^{\dagger}),
    \label{Appendix:multilevel dispersive}
\end{multline}
where $\delta'_n$ is the dispersively shifted energy levels of the transmon in the drive frame, $\delta'_r$ is the dispersively shifted frequency of the resonator, and $\chi_{i,i+1}$ are the dispersive frequency shifts between neighbouring transmon states. Here we assume that the residual drive on the dressed resonator arising from the dispersive rotation can be neglected. 

For the subspace with $r$ excitations in the resonator, we can replace $\ad \hat{a} \rightarrow r$ and write the transmon part of the Hamiltonian as
\begin{equation}
    \frac{\hat{H}_{q,r}}{\hbar} = \sum_n \delta_{n,r} \overline{\op{n_{r}}{n_{r}}} + \Omega_d (\hat{b} + b^{\dagger}),
\end{equation}
where $\overline{\ket{n_{r}}} = \overline{\ket{n,r}}$ and the resonator-induced shifts have been absorbed into an effective energy $\delta_{n,r}$ for the level $\overline{\ket{n_{r}}}$. 

To simulate the experimental results presented here, we consider the case of $\omega_{{d}}$ near-resonant to $\omega_{ge,r}'=\omega_{01,r}'$ while being far-detuned from $\omega_{ef,r}'=\omega_{12,r}'$. Also, for the sake of this calculation, we assume  $\Omega_d \ll \alpha$, the anharmonicity. Hence, the effective contribution of off-resonant driving can be studied perturbatively by writing $\hat{H}_{q,r}$ as
\begin{equation}
\hat{H}_{q,r} = \hat{H}_{q,r}^0 + \hat{H}_{q,r}^1,
\end{equation}
where
\begin{align}
    \hat{H}_{q,r}^0 &= \sum_{n} \delta_{n,r} \overline{\op{n_{r}}{n_{r}}} + \Omega_d (\overline{\op{0_{r}}{1_{r}}} + \overline{\op{1_{r}}{0_{r}}})
\end{align}
is the unperturbed Hamiltonian and 
\begin{align}
\hat{H}_{q,r}^1 &=   \Omega_d \sum_{n>1} \sqrt{n} (\overline{\op{n_{r}}{(n-1)_{r}}} + \overline{\op{(n-1)_{r}}{n_{r}}})
\end{align}
is the perturbation. As a result of the coupler drive, the first two eigenstates of the unperturbed Hamiltonian are given by 
\begin{equation}
    \mqty(\,\overline{\ket{+_{r}}} \,\\ \\ \overline{\ket{-_{r}}}\,) = R_{\frac{\theta_{r}}{2}} \mqty(\,\overline{\ket{0_{r}}} \,\\ \\ \overline{\ket{1_{r}}}\,)
\end{equation}
where $\tan\theta_{r} = 2\Omega_d/\delta_{1,r}$, while the other eigenstates are $\overline{\ket{n_{r}}}$ for $n>1$. Since the drive mixes only adjacent levels, the energy corrections $\hbar \Delta \omega_{\pm,r}$ arising from second-order perturbation theory to the $\overline{\ket{+_{r}}},\overline{\ket{-_{r}}}$ energies $\hbar\omega_{\pm, r}^0$ are only due to the $\overline{\ket{2_{r}}}$ state. These are given by
\begin{align}
    \omega_{\pm,r}^0 &= \frac{\delta_{1,r}}{2} \pm \sqrt{\left(\frac{\delta_{1,r}}{2}\right)^2 + \Omega_d^2},\\
    \Delta \omega_{-,r} &= 2\Omega_d^2 \frac{\sin^2(\theta_{r}/2)}{\omega_{-,r}^0 - \delta_{2,r}},\\
    \Delta \omega_{+,r} &= 2\Omega_d^2 \frac{\cos^2(\theta_{r}/2)}{\omega_{+,r} ^0- \delta_{2,r}}.
\end{align}

We can now determine the transition frequencies for the $\ket{1_p}\rightarrow\ket{3_p}$ ($\omega_{13,p}$) and $\ket{2_p}\rightarrow\ket{4_p}$ ($\omega_{24,p}$) polariton transitions. In the dispersive regime, the polariton states are given by the mapping $\ket{1_p} \equiv \overline{\ket{-,0}}$, $\ket{2_p} \equiv \overline{\ket{+,0}}$, $\ket{3_p} \equiv \overline{\ket{-,1}}$, and $\ket{4_p} \equiv \overline{\ket{+,1}}$. When $\omega_{{d}} = \omega_{ge,0}'$ (as in Fig.~\ref{fig: experimental data - crossplot}), we have $\delta_{1,0} = 0, \delta_{1,1}=-2\chi$. Furthermore, we can also approximate $\delta_{2,0}\approx \delta_{2,1} \approx \alpha \gg \omega_{\pm, r}$. Then, we obtain 
\begin{align}
    \omega_{13,p} = \Tilde{\omega}_r - (\sqrt{\chi^2 + \Omega_d^2} - \Omega_d) + \frac{\Omega_d^2 \cos\theta_1}{\alpha},\\
    \omega_{24,p} = \Tilde{\omega}_r + (\sqrt{\chi^2 + \Omega_d^2} - \Omega_d) - \frac{\Omega_d^2 \cos\theta_1}{\alpha},
\end{align}
which correspond to Eq.~(\ref{eqn:transition_freqs}) discussed in the paper.

\input{refs_polariton.bbl}
\end{document}

%% file: refs_polariton.bbl
%